\documentclass[english]{article}
\usepackage{amssymb}
\usepackage{fontenc}
\usepackage{amsmath}
\usepackage{fontenc}
\usepackage{fontenc}
\usepackage{graphicx}
\usepackage{babel}
\usepackage[latin1]{inputenc}

\input{tcilatex}
\begin{document}

\title{\textbf{Density matrix of a quantum field in a particle-creating
background}}
\author{S.P. Gavrilov\thanks{%
On leave from Department of general and experimental physics, Herzen State
Pedagogical University of Russia, Moyka emb. 48, 191186 St. Petersburg,
Russia; electronic address: gavrilovsergeyp@yahoo.com}, D.M. Gitman\thanks{%
Electronic address: gitman@dfn.if.usp.br}, and J.L. Tomazelli\thanks{%
Dept. Física, CCFM, Universidade Federal de Santa Catarina, CP 476, CEP
88010-970, Santa Catarina, SC, Brazil; electronic address:
tomazelli@fsc.ufsc.br} \\
\\
Instituto de Física, Universidade de São Paulo,\\
Caixa Postal 66318, CEP 05315-970 São Paulo, SP, Brazil}
\maketitle

\begin{abstract}
We examine the time evolution of a quantized field in external backgrounds
that violate the stability of vacuum (particle-creating backgrounds). Our
purpose is to study the exact form of the final quantum state (the density
operator at the final instant of time) that has emerged from a given
arbitrary initial state (from a given arbitrary density operator at the
initial time instant) in the course of evolution. We find a generating
functional that allows one to obtain density operators for an arbitrary
initial state. Averaging over states of the subsystem of antiparticles
(particles), we obtain explicit forms of reduced density operators for the
subsystem of particles (antiparticles). Analyzing one-particle correlation
functions, we establish a one-to-one correspondence between these functions
and the reduced density operators. It is shown that in the general case a 
\emph{\ }presence of bosons (e.g., gluons) in the initial state increases
the creation rate of the same type of bosons. We discuss the question (and
its relation to the initial stage of quark-gluon plasma formation) whether a
thermal form of one-particle distribution can appear even if the final state
of the complete system is not in thermal equilibrium. In this respect, we
discuss some cases when pair-creation by an electric-like field can mimic
the one-particle thermal distribution. We apply our technics to some QFT
problems in slowly varying electric-like backgrounds: electric, SU(3)
chromoelectric, and metric. In particular, we analyze the time and
temperature behavior of the mean numbers of created particles, provided that
the effects of switching the external field on and off are negligible. It is
demonstrated that at high temperatures and in slowly varying electric fields
the rate of particle-creation is essentially time-dependent.

Keywords: external particle-creating field; evolution of an arbitrary
initial state; creation rate; electric, SU(3) chromoelectric and metric
fields; initial thermal distribution.
\end{abstract}

\section{Introduction}

The effect of particle creation from vacuum by an external background
(vacuum instability in external fields) ranks among the most intriguing
nonlinear phenomena in quantum theory. Its theoretical analysis must be
nonperturbative, and its observation in experiment is to verify the
applicability of a theory in the domain of superstrong fields. The study of
this effect started in connection with the so-called Klein \cite{b1}
paradox, and was carried on by Schwinger \cite{S51}, who calculated the
vacuum-to-vacuum transition probability in a constant electric field. A
complete study of particle creation from vacuum by a constant electric field
is presented in \cite{Nikis79,GavG96a}. It should be noted that the effect
can actually be observed as soon as the external field strength approaches
the characteristic value (critical field) $E_{c}=m^{2}c^{3}/|e|\hbar \simeq
1,3\cdot 10^{16}\;V/cm$. Although an actual possibility of creating these
fields under laboratory conditions does not exist at present, $e^{+}e^{-}$%
-pair production by a slowly varying (external) electric field from vacuum
is possibly relevant to phenomenology with the advent of a new laser
technology, which may access the truly strong-field domain.
Electron-positron pairs can also be created by perfectly perturbative
processes in crossed laser beams, when there occurs the merging of several
photons. The control over the dependence between the power and frequency of
lasers makes in possible to determine which of these mechanisms is
responsible for pair-creation.{\large \ }This topic is widely discussed \cite%
{Dre02} at SLAC and TESLA X-ray laser facilities. Such strong fields may be
essential in astrophysics, where characteristic values of the
electromagnetic and gravitational fields in the vicinity of black holes are
enormous. Electric fields near a cosmic string can become extremely strong 
\cite{NCV99}. In this respect, one needs to mention that the Coulomb field
of superheavy nuclei may create electron-positron pairs; see \cite{GMR85}.
Apart from purely QED problems, there arise problems of QFT in which vacuum
instability in various external backgrounds plays an important role, for
example, phase transitions in non-Abelian theories, the problem of boundary
conditions, or the influence of topology on vacuum, the problem of a
consistent vacuum construction in QCD and GUT, multiple particle-creation
within the context of heavy-ion collisions, and so on. Particle creation by
background metrics is important in black-hole physics \cite{Haw75,FroN89}
and also in the study of the dynamics of the early Universe \cite{Grib}.
Recently, it has also been recognized that the presence of a background
electric field must be taken into account in string theory constructions;
see, e.g., \cite{string} and references therein.

Considerable attention has been recently focused on a non-perturbative
parton production from vacuum by a classical chromoelectric field of SU(3) 
\cite{NayN05,GelKL06} and SU(2) \cite{KhaLT06}, in the framework of a modern
version of the known chromoelectric flux tube model \cite{CasNN79}, the
latter being an effective model for the confinement of quarks in QCD
(previously pair-creation by a constant field has been calculated for SU(2)
in \cite{BatMS77} and for SU(3) in \cite{YilC80}). The model ensures a very
good description of the phenomenology of hadron jets in high-energy $%
e^{+}-e^{-}$ and $p-\bar{p}$ collision experiments (a further development of
the basic model and phenomenological applications can be found in the review 
\cite{ManS01}). This model probably describes the initial stage of
quark-gluon plasma formation reasonably well (in particular, the transversal
spectrum of produced soft partons). Such a state may be produced at
high-energy large-hadron colliders such as RHIC (Au-Au collisions at $\sqrt{s%
}=200$ GeV) \cite{McLG05} and LHC (Pb-Pb collisions at $\sqrt{s}=5,5$ TeV) 
\cite{Schu04}. At present, this initial stage is related to an effective
theory, the color glass condensate \cite{McLV94} (see also the review papers 
\cite{IanV03}), which is the coherent limit of quark-gluon plasma at high
energies. In such a picture, after a nuclei collision, a strong classical
chromo-electric-magnetic field is created due to relatively slow
fluctuations of color density. Such a field is sufficiently uniform in the
direction that is transversal to the beam direction and has a longitudinal
chromoelectric component \cite{KhaLT06,KovMW95}. This component is much more
intensive than the transversal component; see \cite{KhaLT06}. Thus, the
color glass condensate picture provides strong arguments in favour of the
chromoelectric flux tube model and allows one to calculate field
configurations in a tube. In particular, such a physical picture allows one
to accept a quasi-constant chromoelectric field as a good approximation at
the above-mentioned initial stage. It should be noted that experimental data
on heavy-ion collisions that exist at present can be interpreted as quantum
parton production by an external chromoelectric field from vacuum and
many-particle states.

There is a considerable interest in particle-creation at finite temperatures
and at a finite particle density, which is basically motivated by heavy-ion
collisions, cosmological QCD phase transitions, and dark matter formation.
For example, thermally-influenced pair-production in constant electric
fields at the one-loop level has been searched for in \cite%
{BuhGF80,HalL95,GanKP95,Gie99}.

The above calculations have been carried out within the theory of a
quantized field placed in an external background. A consistent description
of a complete QED (interacting quantum electromagnetic and matter fields in
a particle-creating background) with unstable vacuum that treats the
interaction with external backgrounds nonperturbatively has been developed
in \cite{Gitman}. Possible generalizations of the formalism to external
gravitational and non-Abelian gauge fields have been presented in \cite%
{BFG81,GavG91}; see also \cite{AmbHN83}. An attempt to extend this technics
to the thermal case has been taken in \cite{GavGF87}. Calculating particle
creation by black-hole metric, Hawking has discovered that the density
matrix of created particles at spatial infinity has a thermal character. The
question arises, is such a character related to the particle-creation
mechanism in general, or to the gravitational origin of the background? A
way to answer this question is to elaborate an adequate technique which
could allow one to include arbitrary mixed initial states, in particular,
thermal initial states, in the corresponding particle creation formalism 
\cite{Gitman}.

In this article, we present a development of the particle-creation formalism 
\cite{Gitman} that is capable to answer some of the above questions. We
examine the time evolution of a quantized field (bosonic or fermionic) in
external backgrounds that violates the stability of vacuum. In fact, we deal
with a quadratic field theory of noninteracting (between themselves)
particles. Our purpose is to study the exact form of the final quantum state
(the density operator at the final instant of time) that has emerged from a
given arbitrary initial state (from a given arbitrary density operator at
the initial time instant) in the course of evolution.

The article is organized as follows. Section 2 has an original, albeit rater
technical character. There, we derive exact expressions for density
operators (more appropriate for the generating density operator) by applying
the path integration method. Some necessary formulas are placed in Appendix.
In Section 3, having an exact expression for the generating density
operator, we obtain reduced density operators for the subsystems of
particles and antiparticles. We introduce and calculate the one-particle
correlation functions and establish a one-to-one correspondence between
these functions and the reduced density operators. In particular, this
allows us to restore the reduced density operator of the complete system
starting from the one-particle distributions (of course, this is possible
only in the model under consideration being a quadratic theory). It is
demonstrated that in the general case the presence of bosons (e.g., gluons)
in the initial state increases the creation of the same kind of bosons. We
discuss in detail the question (and its relation to the initial stage of
quark-gluon plasma formation) whether the thermal form of the one-particle
distribution can appear even if the final state of the complete system is
not in thermal equilibrium.\emph{\ }In Section 4, we discuss the expressions
we obtain for the density operators and one-particle distributions in
electric-like backgrounds: electric, SU(3) chromoelectric, and metric. In
particular, we analyze the density operators and one-particle distributions
in the so-called $T$-constant electric background (a field exists during a
finite period of time $T$), and demonstrate how such a problem is related to
particle creation by gravitational fields (Hawking's effect).\emph{\ }We
present some examples when pair-creation by an electric-like field can mimic
the one-particle thermal distribution. Then, we analyze the time and
temperature behavior of particle-creation when the effects of switching on
and off are negligible. In particular, we demonstrate that at high
temperatures the production rate is non-trivially time-dependent. This
result has to be taken into account at high temperatures.

\section{Density operator in pair-creating backgrounds}

We consider a quantum field $\psi (x)$ in an external background. The
quantum field can be scalar, spinor, etc., and the background can be the
classical electromagnetic, Yang--Mills, or gravitational field. In the
general case, the background is intense, time-dependent, and violates the
stability of vacuum. Such a background must be treated nonperturbatively. We
plan to follow the formalism proposed in \cite{Gitman}.

\subsection{Some relevant relations}

It is assumed that there exists a set of creation and annihilation operators 
$a_{n}^{\dagger }(t_{in})$, $a_{n}(t_{in})$ of particles $a_{n}^{\dagger
}(t_{in})$, $a_{n}(t_{in})$, and antiparticles $b_{n}^{\dagger }(t_{in})$, $%
b_{n}(t_{in})$, respectively, at the initial time instant $t_{in}$ ($%
t_{in}\rightarrow -\infty $), and a set of creation and annihilation
operators of particles, $a_{n}^{\dagger }(t_{out})$, $a_{n}(t_{out})$, and
antiparticles, $b_{n}^{\dagger }(t_{out})$, $b_{n}(t_{out})$, at the final
time instant $t_{out}$ ($t_{out}\rightarrow \infty $). By $n$ we denote a
complete set of possible quantum numbers. The total Hamiltonian $\hat{H}%
\left( t\right) $ of the quantized field under consideration is diagonalized
(and has a canonical form) in terms of the first set at the initial time
instant, and is diagonalized (and has a canonical form) in terms of the
second set at the final time instant. Nonzero (anti)commutators\footnote{%
The subscript $-$ denotes the commutator (in the Bose case), whereas the
subscript $+$ denotes the anticommutator (in the Fermi case)} are given by 
\begin{eqnarray}
&&\,[a_{n}(t_{in}),a_{m}^{\dagger }(t_{in})]_{\pm
}=[a_{n}(t_{out}),a_{m}^{\dagger }(t_{out})]_{\pm }  \notag \\
&&\,=[b_{n}(t_{in}),b_{m}^{\dagger }(t_{in})]_{\pm
}=[b_{n}(t_{out}),b_{m}^{\dagger }(t_{out})]_{\pm }=\delta _{nm}\,.
\label{a.1}
\end{eqnarray}
The vacuum states $|0,t_{in}\rangle $ at $t_{in}$ and $|0,t_{out}\rangle $
at $t_{out}$ are defined as usual: 
\begin{equation*}
a(t_{in})|0,t_{in}\rangle \;=b(t_{in})|0,t_{in}\rangle
=0\,,\;a(t_{out})|0,t_{out}\rangle \;=b(t_{out})|0,t_{out}\rangle =0\,.
\end{equation*}
Therefore, we define these states as the states that minimize the mean value
of the Hamiltonian $\hat{H}\left( t\right) $ at $t=t_{in}$ and $t=t_{out}$,
respectively. According to the general princilples{\large \ }of quantum
theory, the probability amplitude of transiltion from a certain initial
state $b_{m}^{\dagger }(t_{in})\ldots a_{n}^{\dagger
}(t_{in})|0,t_{in}\rangle $ to a certain final state $b_{m^{\prime
}}^{\dagger }(t_{out})\ldots a_{n^{\prime }}^{\dagger
}(t_{out})|0,t_{out}\rangle $ in the Schrödinger picture has the form 
\begin{equation*}
\left\langle 0,t_{out}\left| a_{n^{\prime }}(t_{out})\ldots b_{m^{\prime
}}(t_{out})U\left( t_{out},t_{in}\right) b_{m}^{\dagger }(t_{in})\ldots
a_{n}^{\dagger }(t_{in})\right| 0,t_{in}\right\rangle ,\;
\end{equation*}
where $U\left( t,t^{\prime }\right) $ is a unitary evolution operator of the
system. Let $\hat{\rho}\left( t_{in}\right) =\rho \left( a^{\dagger
}(t_{in}),a(t_{in}),b^{\dagger }(t_{in}),b(t_{in})\right) ,$ $\mathrm{tr\,}%
\hat{\rho}\left( t_{in}\right) =1,$ be the density operator of the system
under consideration at the initial time instant. As the system develops in
time, this density operator becomes $\hat{\rho}\left( t_{out}\right) $ for
antiparticles at the final instant of time: 
\begin{equation}
\hat{\rho}\left( t_{out}\right) =U\left( t_{out},t_{in}\right) \hat{\rho}%
\left( t_{in}\right) U^{\dagger }\left( t_{out},t_{in}\right) \,.
\label{a.4}
\end{equation}

To pass to the Heisenberg picture, we introduce finite-time evolution
operators $\Omega _{\left( \pm \right) },$%
\begin{equation*}
\Omega _{\left( +\right) }=U\left( 0,t_{in}\right) \,,\;\Omega _{\left(
-\right) }=U\left( 0,t_{out}\right) \,,\;U\left( t_{out},t_{in}\right)
=\Omega _{\left( -\right) }^{\dagger }\Omega _{\left( +\right) }\,.
\end{equation*}
We then define a set of creation and annihilation operators $a_{n}^{\dagger
}(in)$, $a_{n}(in)$ of $in$-particles, as well as similar operators $%
b_{n}^{\dagger }(in)$, $b_{n}(in)$ of $in$-antiparticles, a corresponding $%
in $-vacuum $|0,in\rangle ,$ a set of creation and annihilation operators $%
a_{n}^{\dagger }$, $a_{n},$ of $out$-particles and similar operators $%
b_{n}^{\dagger }$, $b_{n}$ of $out$-antiparticles, and a corresponding $out$%
-vacuum $|0\rangle $, 
\begin{eqnarray}
&&\left( a^{\dagger }(in),a(in),b^{\dagger }(in),b(in)\right) =\Omega
_{(+)}\left( a^{\dagger }(t_{in}),a(t_{in}),b^{\dagger
}(t_{in}),b(t_{in})\right) \Omega _{(+)}^{\dagger }\,,  \notag \\
&&\left( a^{\dagger },a,b^{\dagger },b\right) =\Omega _{(-)}\left(
a^{\dagger }(t_{out}),a(t_{out}),b^{\dagger }(t_{out}),b(t_{out})\right)
\Omega _{(-)}^{\dagger }\,,  \notag \\
&&\,|0,in\rangle =\Omega _{(+)}|0,t_{in}\rangle \,,\;|0\rangle =\Omega
_{(-)}|0,t_{out}\rangle \,.  \label{a.2}
\end{eqnarray}
The $in$- and $out$-operators obey the canonical commutation relations (\ref%
{a.1}).

The entire information concerning the processes of particle creation,
annihilation and scattering is contained in the elementary probability
amplitudes 
\begin{eqnarray}
&&w\left( +|+\right) _{mn}=c_{v}^{-1}\langle 0\left| a_{m}a_{n}^{\dagger
}(in)\right| 0,in\rangle ,  \notag \\
&&w\left( -|-\right) _{nm}=c_{v}^{-1}\langle 0\left| b_{m}b_{n}^{\dagger
}(in)\right| 0,in\rangle \,,  \notag \\
&&w\left( 0|-+\right) _{nm}=c_{v}^{-1}\langle 0\left| b_{n}^{\dagger
}(in)a_{m}^{\dagger }(in)\right| 0,in\rangle \,,  \notag \\
&&w\left( +-|0\right) _{mn}=c_{v}^{-1}\langle 0\left| a_{m}b_{n}\right|
0,in\rangle \,,  \label{a.2a}
\end{eqnarray}
where $c_{v}$ is the vacuum-to-vacuum transition amplitude, $c_{v}=\langle
0|0,in\rangle \,$.

The sets of $in$ and $out$-operators are related to each other by a linear
canonical transformation (it is sometimes called the Bogolyubov
transformation). As has been demonstrated in the general case, such a
relation has the form (see \cite{Gitman}) 
\begin{eqnarray}
&&V\left( a^{\dagger },a,b^{\dagger },b\right) V^{\dagger }=\left(
a^{\dagger }(in),a(in),b^{\dagger }(in),b(in)\right) \,,  \notag \\
&&\,|0,in\rangle =V|0\rangle \,\;\left( c_{v}=\langle 0|V|0\rangle \right)
\,,  \label{a.3}
\end{eqnarray}%
where a unitary operator $V$ has the form 
\begin{equation}
V=v_{4}v_{3}v_{2}v_{1}\,,  \label{a.3a}
\end{equation}%
and\footnote{%
We use condenced notation, for example, 
\begin{equation*}
bw\left( 0|-+\right) a=\sum_{n,m}b_{n}w\left( 0|-+\right) _{nm}a_{m}\,.
\end{equation*}%
} 
\begin{eqnarray}
&&v_{1}=\exp \left\{ -\kappa bw\left( 0|-+\right) a\right\} \,,\;v_{2}=\exp
\left\{ a^{\dagger }\ln w\left( +|+\right) a\right\} \,,  \notag \\
&&v_{3}=\exp \left\{ -\kappa b\ln w\left( -|-\right) b^{\dagger }\right\}
\,,\;v_{4}=\exp \left\{ -\kappa a^{\dagger }w\left( +-|0\right) b^{\dagger
}\right\} \,,  \notag \\
&&\kappa =\left\{ 
\begin{array}{l}
1\mathrm{\;\ Fermi\;case} \\ 
-1\;\ \mathrm{Bose\;case}%
\end{array}%
\right. .  \label{a.3b}
\end{eqnarray}%
Using this expression for $V$, one can find 
\begin{equation}
c_{v}=\langle 0|V|0\rangle =\exp \left\{ -\kappa \mathrm{tr}\ln w\left(
-|-\right) \right\} \,.  \label{a.3c}
\end{equation}

The density operator $\check{\rho}$ of the system under consideration in the
Heisenberg picture is defined as 
\begin{eqnarray}
\check{\rho} &=&\Omega _{(+)}\hat{\rho}\left( t_{in}\right) \Omega
_{(+)}^{\dagger }\,=\rho \left( a^{\dagger }(in),a(in),b^{\dagger
}(in),b(in)\right) \,,  \notag \\
\check{\rho} &=&\Omega _{(-)}\hat{\rho}\left( t_{out}\right) \Omega
_{(-)}^{\dagger }\,,\;\mathrm{tr\,}\check{\rho}=1\,.  \label{a.5a}
\end{eqnarray}

Suppose a physical quantity is given by an operator $\hat{F}\left(
t_{out}\right) $ at the final time instant as 
\begin{equation}
\hat{F}\left( t_{out}\right) =F\left( a^{\dagger
}(t_{out}),a(t_{out}),b^{\dagger }(t_{out}),b(t_{out})\right) \,.
\label{a.6}
\end{equation}
Then its mean value at the final instant of time is given by 
\begin{equation}
\langle F\rangle =\mathrm{tr\,}\left[ \hat{F}\left( t_{out}\right) \hat{\rho}%
\left( t_{out}\right) \right] =\mathrm{tr}\left[ \mathrm{\,}\check{F}\check{%
\rho}\right] \,,  \label{a.7}
\end{equation}
where 
\begin{equation}
\check{F}=\Omega _{(-)}\hat{F}\left( t_{out}\right) \Omega _{(-)}^{\dagger
}=F\left( a^{\dagger },a,b^{\dagger },b\right)  \label{a.8}
\end{equation}
is the operator of a physical quantity in the Heisenberg representation%
\footnote{%
All operators in the Heisenberg representation are denoted by the turned
over hat in what follows, e.g. $\check{A}.$}.

The amplitudes (\ref{a.2a}) can be calculated with the help of some
appropriate sets of solutions of the corresponding relativistic wave
equation (RWE) with an external field (Klein--Gordon, Dirac, linearized
Yang--Mills), see \cite{Gitman,GavG91}, as follows. Namely, the $in$%
-particles are associated with a complete set ($in$-set) of solutions of the
RWE $\left\{ _{\zeta }\psi _{n}(x)\right\} $ with asymptotics $_{\zeta }\psi
_{n}(t_{in},\mathbf{x})$ at the initial time-instant $t_{in}$ being
eigenvectors of the corresponding one-particle Hamiltonian $\mathcal{H}(t)$, 
\begin{equation}
\mathcal{H}(t_{in})_{\zeta }\psi _{n}(t_{in},\mathbf{x})=\zeta \varepsilon
_{n}^{(\zeta )}{}_{\zeta }\psi _{n}(t_{in},\mathbf{x})\,,  \label{asy1}
\end{equation}
where $\varepsilon _{n}^{(\zeta )}$\ are the energies of $in$-particles in a
state specified by a complete set of quantum numbers $n$, and $\varepsilon
_{n}^{(\pm )}>0$. Here and elsewhere, $\zeta =\pm \,$, being $(+)$ for
particles and $(-)$ for antiparticles. For the sake of simplicity of
exposition, we use such a gauge-fixing of the external field that condition (%
\ref{asy1}) should{\large \ }have the same form as it has in the absence of
the external field.{\large \ }The $out$-particles are associated with the
complete $out$-set of solutions $\left\{ ^{\zeta }\psi _{n}\left( x\right)
\right\} $ of the RWE with asymptotics $^{\zeta }\psi _{n}(t_{out},\mathbf{x}%
)$ at $t_{out}$ being eigenvectors of the Hamiltonian $\mathcal{H}(t)$ at $%
t_{out}$, namely, 
\begin{equation}
\mathcal{H}(t_{out})\,^{\zeta }\psi _{n}(t_{out},\mathbf{x})=\zeta \tilde{%
\varepsilon}_{n}^{(\zeta )}\,^{\zeta }\psi _{n}(t_{out},\mathbf{x})\,,
\label{asy2}
\end{equation}
where $\tilde{\varepsilon}_{n}^{(\pm )}$ are the energies of $out$-particles
in a state specified by a complete set of quantum numbers $n$, and $\tilde{%
\varepsilon}_{n}^{(\pm )}>0$. We suppose that the external field is such
that soluitons of this eigevalue problem actually exist.

The $out$-set can be decomposed in the $in$-set as follows: 
\begin{equation}
{}^{\zeta }\psi (x)={}_{+}\psi (x)G\left( {}_{+}|{}^{\zeta }\right)
+{}\kappa _{-}\psi (x)G\left( {}_{-}|{}^{\zeta }\right) \,,  \label{emt18}
\end{equation}
where $\kappa =+1$ for fermions and $\kappa =-1$ for bosons. The
decomposition coefficients $G\left( {}_{\zeta }|{}^{\zeta ^{\prime }}\right) 
$ are expressed via inner products of these sets, $G\left( {}_{\zeta
}|{}^{\zeta ^{\prime }}\right) _{mn}=\left( _{\zeta }\psi _{m},^{\zeta
^{\prime }}\psi _{n}\right) $. These coefficients obey the unitarity
relations 
\begin{eqnarray}
&&G\left( ^{\zeta }|_{+}\right) G\left( {}_{+}|{}^{\zeta }\right) +\kappa
G\left( ^{\zeta }|_{-}\right) G\left( {}_{-}|{}^{\zeta }\right) =\zeta ^{%
\frac{1-\kappa }{2}}\,,  \notag \\
&&G\left( {}_{\zeta }|{}^{+}\right) G\left( ^{+}|_{\zeta }\right) +\kappa
G\left( {}_{\zeta }|{}^{-}\right) G\left( ^{-}|_{\zeta }\right) =\zeta ^{%
\frac{1-\kappa }{2}}\,,  \notag \\
&&G\left( {}_{+}|{}^{+}\right) G\left( ^{+}|_{-}\right) +\kappa G\left(
{}_{+}|{}^{-}\right) G\left( ^{-}|_{-}\right) =0\,,  \notag \\
&&G\left( ^{+}|_{+}\right) G\left( {}_{+}|{}^{-}\right) +\kappa G\left(
^{+}|_{-}\right) G\left( {}_{-}|{}^{-}\right) =0\,,  \label{vst12}
\end{eqnarray}
that follow from the normalization conditions for the solutions. Here, the
notation $G\left( ^{\zeta ^{\prime }}|_{\zeta }\right) =G\left( _{\zeta
}|^{\zeta ^{\prime }}\right) ^{\dagger }$ has been used.

The quantum Heisenberg field $\check{\psi}(x)$ can be expressed both from
the creation and annihilation operators of $in$-particles and from the
creation and annihilation operators of $out$-particles: 
\begin{equation*}
\check{\psi}(x)={}_{+}\psi (x)a(in)+{}_{-}\psi (x)b^{\dagger
}(in)\,={}^{+}\psi (x)a+{}^{-}\psi (x)b^{\dagger }.
\end{equation*}%
This makes it possible to express the linear canonical transformation
between the sets of $in$ and $out$-operators in terms of the coefficients $%
G\left( ^{\zeta ^{\prime }}|_{\zeta }\right) $, 
\begin{eqnarray}
a &=&G\left( ^{+}|_{+}\right) a(in)+G\left( ^{+}|_{-}\right) b^{\dagger
}(in),  \notag \\
\kappa b^{\dagger } &=&G\left( ^{-}|_{+}\right) a(in)+G\left(
^{-}|_{-}\right) b^{\dagger }(in),  \label{a.9}
\end{eqnarray}%
and their Hermitian conjugated ones. Using relations (\ref{a.9}), one finds
the amplitudes (\ref{a.2a}) in the form: 
\begin{eqnarray}
&&w\left( +|+\right) =G\left( _{+}|^{+}\right) ^{-1}\;,\;\;w\left(
-|-\right) =\kappa G\left( ^{-}|_{-}\right) ^{-1}\;,  \notag \\
&&w\left( 0|-+\right) =-G\left( ^{-}|_{-}\right) ^{-1}G\left(
^{-}|_{+}\right) =\kappa G\left( _{-}|^{+}\right) G\left( _{+}|^{+}\right)
^{-1}\;,  \notag \\
&&w\left( +-|0\right) =\kappa G\left( _{+}|^{+}\right) ^{-1}G\left(
_{+}|^{-}\right) =-G\left( ^{+}|_{-}\right) G\left( ^{-}|_{-}\right) ^{-1}\;.
\label{a.10}
\end{eqnarray}%
Using the same relations, one finds that the differential mean numbers $%
\aleph _{m}^{(\zeta )}$ of particles/antiparticle created from vacuum are 
\begin{eqnarray}
\aleph _{m}^{(+)} &=&\left\langle 0,in\left\vert a_{m}^{\dagger
}a_{m}\right\vert 0,in\right\rangle =\left[ G\left( ^{+}|_{-}\right) G\left(
_{-}|^{+}\right) \right] _{mm}\,,\;  \notag \\
\aleph _{m}^{(-)} &=&\left\langle 0,in\left\vert b_{m}^{\dagger
}b_{m}\right\vert 0,in\right\rangle =\left[ G\left( ^{-}|_{+}\right) G\left(
_{+}|^{-}\right) \right] _{mm}\,.  \label{dm27}
\end{eqnarray}%
Using the unitarity relations (\ref{vst12}) and representation (\ref{a.3c}),
we find that the probability $P^{v}$ of a vacuum to remain a vacuum is given
by 
\begin{equation}
P^{v}=|c_{v}|^{2}=\exp \left\{ \kappa \mathrm{tr}\ln \left[ 1-\kappa G\left(
^{-}|_{+}\right) G\left( {}_{+}|{}^{-}\right) \right] \right\} \,.
\label{a.11}
\end{equation}

Therefore, we can see that the non-stability of vacuum in external fields is
mainfest in the fact that $P^{v}<1$, which is in one-to-one-correspondence
with the fact that the differential mean numbers $\aleph _{m}^{(\zeta )}$
are different from zero. We can now establish the following general
property. Let the time-dependent external field turn on at the time instant $%
t_{1}$ and turn off at the time instant $t_{2}$, and in general behaves
arbitrarily, on condition that the solutions of problems (\ref{asy1}) and (%
\ref{asy2}) do exist. Then, the $in$-set of solutions $\left\{ _{\zeta }\psi
_{n}(x)\right\} $ is defined equally for any solution of the time instant $%
t_{in}<t_{1}$, and the $out$-set of solutions $\left\{ ^{\zeta }\psi
_{n}\left( x\right) \right\} $ is defined equally for any choice of the time
instant $t_{out}>t_{2}$. Therefore, the coefficients $G\left( ^{\zeta
^{\prime }}|_{\zeta }\right) $, and consequently all the above-calculated
amplitudes, depend only on the dependence of the external field at the
finite interval of time between $t_{1}$ and $t_{2}$. This also implies that
the $in$- and $out$- vacua, as well as the $in$- and $out$-operators\ of
creation and annihilation, do not depend on the choice of the time instant $%
t_{in}$ and $t_{out}$, respectively, if $t_{in}<t_{1}$ and $t_{out}>t_{2}$.

\subsection{Generating density operator}

We introduce the following generating operator $\check{R}\left( J\right) $: 
\begin{eqnarray}
&&\check{R}\left( J\right) =\frac{1}{Z}\underline{\check{R}}\left( J\right)
\,,\;\;\mathrm{tr}\check{R}\left( J\right) =1\,,  \label{dm1} \\
&&\underline{\check{R}}\left( J\right) =\mathcal{N}_{in}\exp \left[
a^{\dagger }(in)\left( \mathbb{J}^{(+)}-1\right) a(in)+b^{\dagger
}(in)\left( \mathbb{J}^{(-)}-1\right) b(in)\right] \,,  \notag
\end{eqnarray}
where the Grassmann-even variables $J=\left( J_{n}^{(\zeta )}\right) $ are
sources; $\mathbb{J}_{mn}^{(\zeta )}=\delta _{mn}J_{n}^{(\zeta )}$; $%
\mathcal{N}_{in}$ is the sign of the normal form with respect to the $in$%
-vacuum, and $Z=\mathrm{tr}\underline{\check{R}}\left( J\right) $ is a
normalization factor (statistical sum).\ 

In order to complete the calculation, it is efficient to use the path
integral representation. In the fermion case, we use a path integral over
anticommuting (Grassmann) variables, which is understood as Berezin's
integral \cite{Ber65} at $\kappa =1$, 
\begin{equation}
:e^{-\kappa a^{\dagger }Ka}:=\det K^{\kappa }:\int \exp \left\{ \kappa
\lambda ^{\ast }K^{-1}\lambda +a^{\dagger }\lambda +\lambda ^{\ast
}a\right\} \Pi d\lambda ^{\ast }d\lambda :,  \label{ap8}
\end{equation}
where $a^{\dagger }$, $a$ are some creation and annihilation operators, and $%
:\ldots :$ realizes the normal form of the operators $a^{\dagger }$, $a$.
All operators $a^{\dagger }$ and $a$ can be considered as Grassmann-odd
variables under the normal form; therefore, we can calculate the complete
path integral (\ref{ap8}) as a Gaussian integral over Grassmann-odd
variables. In the boson case, we use the path integral\ (\ref{ap8}) over
commuting variables at $\kappa =-1$. In this case, we can regard all the
operators $a^{\dagger }$ and $a$ as bosonic (ordinary) variables under the
normal form sign, so that the path integral in\ (\ref{ap8}) is a usual
Gaussian path integral, where $\lambda ^{\ast }K^{-1}\lambda >0$.

Note that the trace of the normal product of creation and annihilation
operators can be calculated by using the path integral representation,
according to (\ref{ap12}), see Appendix. For example, by calculating $Z$ we
obtain 
\begin{equation}
Z=\exp \left\{ \kappa \sum_{n}\left[ \ln \left( 1+\kappa \mathbb{J}%
^{(+)}\right) \right] _{nn}+\kappa \sum_{m}\left[ \ln \left( 1+\kappa 
\mathbb{J}^{(-)}\right) \right] _{mm}\right\} \,.  \label{dm1c}
\end{equation}

Having at our disposal the generating operator (\ref{dm1}), we can obtain
the different density operators (in the Heisenberg representation)
corresponding to different initial states of the system. We represent some
examples below:

a) By setting $J=0,$ we obtain a density operator $\check{\rho}_{v}$ of the
system that is found in the pure vacuum state at the initial time instant, 
\begin{equation*}
\check{\rho}_{v}=\check{R}\left( 0\right) \,.
\end{equation*}
Indeed, using relation (\ref{f3}) from Appendix, we obtain 
\begin{equation}
\check{\rho}_{v}=\mathcal{N}_{in}\exp \left\{ -\left[ a^{\dagger
}(in)a(in)+b^{\dagger }(in)b(in)\right] \right\} =|0,in\rangle \langle
0,in|\,.  \label{dm1a}
\end{equation}

In addition, we define the following generating functional of moments: 
\begin{eqnarray}
&&\Phi ^{v}\left( J\right) =\langle 0,in|\,\exp \left[ a^{\dagger }\mathbb{J}%
^{(+)}a+b^{\dagger }\mathbb{J}^{(-)}b\right] |0,in\rangle =\mathrm{tr\,}%
\check{\phi}\left( J\right) \,,  \notag \\
&&\check{\phi}\left( J\right) =\exp \left[ a^{\dagger }\mathbb{J}%
^{(+)}a+b^{\dagger }\mathbb{J}^{(-)}b\right] \check{\rho}_{v}\,,
\label{jfm1}
\end{eqnarray}
which is useful to investigate the final state evolved from vacuum at the
initial time instant.

b) The density operator $\check{\rho}_{\{m\}_{M};\{n\}_{N}}$ of the system
which is found in a pure state with $M$ particles and $N$ antiparticles
(with the quantum numbers $\left\{ m_{1},\ldots ,m_{M}\right\} =\{m\}_{M}$
and $\left\{ n_{1},\ldots ,n_{N}\right\} =\{n\}_{N}$, respectively) at the
initial time instant can be obtained from the generating operator $%
\underline{\check{R}}\left( J\right) $ as follows: 
\begin{equation}
\check{\rho}_{\{m\}_{M};\{n\}_{N}}=\left. \frac{\partial ^{M+N}\underline{%
\check{R}}\left( J\right) }{\partial (J_{m_{1}}^{(+)}\ldots
J_{m_{M}}^{(+)}J_{n_{1}}^{(-)}\ldots J_{n_{N}}^{(-)})}\right| _{J=0}=|\Psi
_{\{m\}_{M};\{n\}_{N}}\left( in\right) \rangle \langle \Psi
_{\{m\}_{M};\{n\}_{N}}\left( in\right) |\,,  \label{dm2a}
\end{equation}
where 
\begin{eqnarray*}
&&\,|\Psi _{\{m\}_{M};\{n\}_{N}}\left( in\right) \rangle
=\prod_{i=1}^{M}a_{m_{i}}^{\dagger }(in)\prod_{j=1}^{N}b_{n_{j}}^{\dagger
}(in)|0,in\rangle \,, \\
&&\left\langle \Psi _{\{m\}_{M};\{n\}_{N}}\left( in\right) \right|
=\left\langle 0,in\right|
\prod_{j=1}^{N}b_{n_{j}}(in)\prod_{i=1}^{M}a_{m_{i}}(in)\,.
\end{eqnarray*}

c) Let us set 
\begin{equation}
J_{n}^{(\zeta )}=e^{-E_{n}^{(\zeta )}},\;E_{n}^{(\zeta )}=\beta \left(
\varepsilon _{n}^{(\zeta )}-\mu ^{(\zeta )}\right) \,,\;\beta ^{-1}=\Theta
\,,  \label{dm3}
\end{equation}
where $\varepsilon _{n}^{(\zeta )}$ are the energies of particles or
antiparticles with the quantum numbers $n$; $\mu ^{(\zeta )}$ are the
corresponding chemical potentials, and $\Theta $ is the absolute
temperature. One can see that with such a choice of sources the generating
operator (\ref{dm1}) becomes the density operator $\check{\rho}_{\beta }$ of
the system that has been in thermal equilibrium at the initial time instant.
Using relation (\ref{f4}) from Appendix, we obtain an explicit expression
for $\check{\rho}_{\beta }$, namely, 
\begin{eqnarray*}
&&\check{\rho}_{\beta }=\check{R}\left( e^{-E_{n}^{(\zeta )}}\right) =\frac{1%
}{Z}\exp \left\{ -\left[ a^{\dagger }(in)E^{(+)}a(in)+b^{\dagger
}(in)E^{(-)}b(in)\right] \right\} \,, \\
&&Z=\exp \left\{ \kappa \sum_{n}\ln \left( 1+\kappa e^{-E_{n}^{(+)}}\right)
+\kappa \sum_{m}\ln \left( 1+\kappa e^{-E_{n}^{(-)}}\right) \right\} \,,
\end{eqnarray*}
or 
\begin{equation}
\,\check{\rho}_{\beta }=Z^{-1}\exp \left\{ -\beta \left[ \check{H}%
-\sum_{\zeta =\pm }\mu ^{(\zeta )}\check{N}^{(\zeta )}\right] \right\} \,,
\label{dm4}
\end{equation}
where $\check{H}$ is the Hamiltonian of the system (written in terms of $in$%
-operators); $\check{N}^{(\zeta )}$ are the operators of $in$-particle or $%
in $-antiparticle numbers, 
\begin{eqnarray*}
&&\check{H}=a^{\dagger }(in)\varepsilon ^{(+)}a(in)+b^{\dagger
}(in)\varepsilon ^{(-)}b(in)\,, \\
&&\check{N}^{(+)}=a^{\dagger }(in)a(in)\,,\;\check{N}^{(-)}=b^{\dagger
}(in)b(in)\,,
\end{eqnarray*}
and the matrices $E^{(\zeta )}$ and $\varepsilon ^{(\zeta )}$ are defines
as\ $E_{mn}^{(\zeta )}=\delta _{mn}E_{n}^{(\zeta )}$ and $\varepsilon
_{mn}^{(\zeta )}=\delta _{mn}\varepsilon _{n}^{(\zeta )}$.

We can see that the problem of calculating the mean value of an operator $%
\hat{F}\left( t_{out}\right) $ for the system being at the final time
instant is related to the problem of calculating the quantity $\mathrm{tr\,}%
\left[ \check{F}\underline{\check{R}}\left( J\right) \right] ,$ where $%
\check{F}$\ is a Heisenberg operator corresponding to $\hat{F}\left(
t_{out}\right) $. Such a quantity can be represented as follows: 
\begin{eqnarray}
&&\mathrm{tr\,}\left[ \check{F}\underline{\check{R}}\left( J\right) \right]
=\sum_{M,N=0}^{\infty }\sum_{\{m\}\left\{ n\right\} }\frac{1}{M!N!}\langle
\Psi \left( \left\{ m\right\} _{M},\left\{ n\right\} _{N}\right) |\check{F}%
\underline{\check{R}}\left( J\right) |\Psi \left( \left\{ m\right\}
_{M},\left\{ n\right\} _{N}\right) \rangle \,,  \notag \\
&&|\Psi \left( \left\{ m\right\} _{M},\left\{ n\right\} _{N}\right) \rangle
=a_{m_{1}}^{\dagger }\ldots a_{m_{M}}^{\dagger }b_{n_{1}}^{\dagger }\ldots
b_{n_{N}}^{\dagger }|0\rangle \,,  \notag \\
&&\left\langle \Psi \left( \left\{ m\right\} _{M},\left\{ n\right\}
_{N}\right) \right| =\left\langle 0\right| b_{n_{N}}\ldots
b_{n_{1}}a_{m_{M}}\ldots a_{m_{1}}  \label{dm5}
\end{eqnarray}
Calculating $\mathrm{tr\,}\left[ \check{F}\underline{\check{R}}\left(
J\right) \right] $ according to (\ref{dm5}), it is convenient to have an
expression for the operator $\underline{\check{R}}\left( J\right) $ in terms
of $out$-operators. One can see that such an expression has the form 
\begin{equation}
\underline{\check{R}}\left( J\right) =VU\left( J\right) V^{\dagger
}\,,\;\;U\left( J\right) =:\exp \left[ a^{\dagger }\left( \mathbb{J}%
^{(+)}-1\right) a+b^{\dagger }\left( \mathbb{J}^{(-)}-1\right) b\right] :\,,
\label{dm8}
\end{equation}
where $:\cdots :$ is the sign of the normal form with respect to the $out$%
-vacuum, and the operator $V$ is defined by (\ref{a.3a}). A normal form of
the operator $\underline{\check{R}}\left( J\right) $ with respect to the $%
out $-vacuum is calculated below.

\subsection{The normal form of the generating operator}

First of all, we rewrite the operator expression (\ref{dm8}) as follows: 
\begin{equation}
\underline{\check{R}}\left( J\right) =v_{4}v_{3}v_{2}\tilde{Y}\left(
J\right) v_{2}^{\dagger }v_{3}^{\dagger }v_{4}^{\dagger }\,,\;\;\tilde{Y}%
=v_{1}U\left( J\right) v_{1}^{\dagger }\,,  \label{dm14a}
\end{equation}%
where the operators $\nu _{i},$ $i=1,...,4,$ are given by (\ref{a.3b}).
Using formula (\ref{f1}) from Appendix, we represent the operator $\tilde{Y}%
(J)$ in the form 
\begin{eqnarray}
&&\tilde{Y}\left( J\right) =Y\left( J\right) U\left( J\right) \,,\;Y\left(
J\right) =\exp \left( -bBa\right) \exp \left( -a^{\dagger }A\left( J\right)
b^{\dagger }\right) \,,  \notag \\
&&A\left( J\right) =\mathbb{J}^{(+)}B^{\dagger }\mathbb{J}^{(-)},\;B=\kappa
w\left( 0|-+\right) \,.  \label{ap7a}
\end{eqnarray}

Both operator exponents in the expression for $Y\left( J\right) $ can be
written in terms of Gaussian path integrals.

Consider, first of all, the fermi-particle case. In this case, we can treat
the anticommuting operators $a$ and $b$ (or $a^{+}$ and $b^{+}$) as
Grassmann-odd variables. Then, according to representation (\ref{ap8}) at $%
\kappa =1$, we have 
\begin{eqnarray}
&&Y=\det A\,\det B\,\int \exp \left( \tilde{\lambda}^{\ast }B^{-1}\tilde{%
\lambda}+\lambda ^{\ast }A^{-1}\lambda \right) \Phi \Pi d\tilde{\lambda}%
^{\ast }d\tilde{\lambda}d\lambda ^{\ast }d\lambda \,,  \notag \\
&&\Phi =\exp \left( b\tilde{\lambda}+\tilde{\lambda}^{\ast }a\right) \exp
\left( a^{\dagger }\lambda +\lambda ^{\ast }b^{\dagger }\right) \,.
\label{ap7b}
\end{eqnarray}
With the help of relation (\ref{dm14b}) from Appendix, we represent the
operator $\Phi $ in the normal form 
\begin{equation*}
\Phi =:\exp \left( a^{\dagger }\lambda +\lambda ^{\ast }b^{\dagger }+b\tilde{%
\lambda}+\tilde{\lambda}^{\ast }a+\tilde{\lambda}^{\ast }\lambda +\tilde{%
\lambda}\lambda ^{\ast }\right) :\,.
\end{equation*}
Then, using formula (\ref{ap8}) we can calculate the complete path integral (%
\ref{ap7b}). In the Bose case, we can examine all the operators $a^{\dagger
} $, $b^{\dagger }$, $a$, and $b$ as bosonic (ordinary) variables under the
normal form sign, so that the operator $Y\left( J\right) $ can be
represented as a usual Gaussian path integral by applying representation (%
\ref{ap8}) at $\kappa =-1$. Calculating these Gaussian integrals, we obtain
the normal form of the operator $Y$, 
\begin{eqnarray}
&&Y=\det \left( 1+\kappa AB\right) ^{\kappa }:\exp \left\{ -a^{\dagger
}A_{++}a-b^{\dagger }A_{--}b-a^{\dagger }A_{+-}b^{\dagger }-bA_{-+}a\right\}
:\,,  \notag \\
&&A_{++}=\kappa AB\left( 1+\kappa AB\right) ^{-1},\;A_{--}^{T}=\kappa
BA\left( 1+\kappa BA\right) ^{-1}\,,  \notag \\
&&A_{+-}=\left( 1+\kappa AB\right) ^{-1}A\,,\;A_{-+}=B\left( 1+\kappa
AB\right) ^{-1}\,.  \label{ap10a}
\end{eqnarray}

Using relation (\ref{dm14c}) from Appendix, we represent the operator $%
\tilde{Y}$ in the normal form 
\begin{eqnarray}
&&\tilde{Y}=\det \left( 1+\kappa AB\right) ^{\kappa }:\exp \left\{
-a^{\dagger }\widetilde{A}_{++}a-b^{\dagger }\widetilde{A}_{--}b-a^{\dagger }%
\widetilde{A}_{+-}b^{\dagger }-b\widetilde{A}_{-+}a\right\} :\;,  \notag \\
&&\widetilde{A}_{++}=1-\left( 1-A_{++}\right) \mathbb{J}^{(+)},\;\widetilde{A%
}_{--}=1-\left( 1-A_{--}\right) \mathbb{J}^{(-)}\,,  \notag \\
&&\widetilde{A}_{+-}=\widetilde{A}_{-+}^{\dagger },\;\widetilde{A}_{-+}=%
\mathbb{J}^{(-)}A_{-+}\mathbb{J}^{(+)}\,.  \label{ap10}
\end{eqnarray}
\ \ \ \ With the help of relation (\ref{a.3c}), we rewrite the operator $%
v_{3}$ as follows: 
\begin{equation*}
v_{3}=\exp \left[ -\kappa b\ln w\left( -|-\right) b^{\dagger }\right]
=c_{v}\exp \left[ b^{\dagger }\ln w\left( -|-\right) ^{T}b\right] \,.
\end{equation*}
Then, using formulas (\ref{f4}) derived in Appendix, we represent the
operators $v_{3}v_{2}$ and $v_{2}^{\dagger }v_{3}^{\dagger }$ from (\ref%
{dm14a}) in the normal form as follows: 
\begin{eqnarray*}
&&v_{3}v_{2}=c_{v}:\exp \left[ b^{\dagger }\left( w\left( -|-\right)
^{T}-1\right) b\right] \exp \left[ a^{\dagger }\left( w\left( +|+\right)
-1\right) a\right] :\,, \\
&&v_{2}^{\dagger }v_{3}^{\dagger }=c_{v}^{\ast }:\exp \left[ a^{\dagger
}\left( w\left( +|+\right) ^{\dagger }-1\right) a\right] \exp \left[
b^{\dagger }\left( w\left( -|-\right) ^{T\dagger }-1\right) b\right] :\,.
\end{eqnarray*}
Finally, using relation (\ref{dm14c}), we obtain the normal form of the
operator $\underline{\check{R}}\left( J\right) ,$%
\begin{eqnarray}
&&\underline{\check{R}}\left( J\right) =\left| c_{v}\right| ^{2}\det \left(
1+\kappa AB\right) ^{\kappa }:\exp \left[ -a^{\dagger }\left( 1-D_{+}\right)
a-b^{\dagger }\left( 1-D_{-}\right) b-a^{\dagger }C^{\dagger }b^{\dagger
}-bCa\right] :,  \notag \\
&&D_{+}=w\left( +|+\right) \left( 1+\kappa AB\right) ^{-1}\mathbb{J}%
^{(+)}w\left( +|+\right) ^{\dagger }\,,  \notag \\
&&D_{-}^{T}=w\left( -|-\right) ^{\dagger }\mathbb{J}^{(-)}\left( 1+\kappa
BA\right) ^{-1}w\left( -|-\right) \,,  \notag \\
&&C=w\left( -|-\right) ^{\dagger }\mathbb{J}^{(-)}B\left( 1+\kappa AB\right)
^{-1}\mathbb{J}^{(+)}w\left( +|+\right) ^{\dagger }+\kappa w\left(
+-|0\right) ^{\dagger }\,.  \label{dm15}
\end{eqnarray}

Representation (\ref{dm15}) is useful, since it allows one to calculate the
trace (\ref{dm5}) by using the path integral techniques described in
Appendix; see, eq. (\ref{ap12}).

As an example, let us consider the density operator $\check{\rho}_{v}$
defined by (\ref{dm1a}). Using (\ref{dm15}), we represent this operator in
terms of $out$-operators, as well as in the normal form 
\begin{equation}
\check{\rho}_{v}=\underline{\check{R}}\left( 0\right) =\left| c_{v}\right|
^{2}:\exp \left[ -a^{\dagger }a-b^{\dagger }b-\kappa a^{\dagger }w\left(
+-|0\right) b^{\dagger }-\kappa bw\left( +-|0\right) ^{\dagger }a\right] :\,.
\label{dm15a}
\end{equation}

In a similar way, the operator $\check{\phi}\left( J\right) $ in the
expression of generating functional (\ref{jfm1}) can be transformed to the
normal form 
\begin{equation}
\check{\phi}\left( J\right) =\left| c_{v}\right| ^{2}:\exp \left[
-a^{\dagger }a-b^{\dagger }b-\kappa a^{\dagger }e^{\mathbb{J}^{(+)}}w\left(
+-|0\right) e^{\mathbb{J}^{(-)}}b^{\dagger }-\kappa bw\left( +-|0\right)
^{\dagger }a\right] :\,,  \label{jfm2}
\end{equation}
where the representation (\ref{dm15a}) for $\check{\rho}_{v}$ has been used.
Then, applying the path integral representation for traces (\ref{ap12}), and
using formula (\ref{dm14c}), we represent the generating functional of
momenta as follows: 
\begin{equation}
\Phi ^{v}\left( J\right) =\left| c_{v}\right| ^{2}\exp \left\{ \kappa 
\mathrm{tr}\ln \left[ 1+\kappa w\left( +-|0\right) ^{\dagger }e^{\mathbb{J}%
^{(+)}}w\left( +-|0\right) e^{\mathbb{J}^{(-)}}\right] \right\} \,.
\label{jfm3}
\end{equation}

\section{Reduced density operators and correlation functions}

\subsection{Reduced density operators}

In the general case, the states of the system under consideration at the
final time instant contain both particles and antiparticles due to the
pair-creation by external fields and the structure of the initial state. On
the other hand, we are often interested in physical quantities $F_{\pm }$
which describe only particles ($+$) or antiparticles ($-$) at the final time
instant. The corresponding operators $\check{F}_{\pm }$ are functions of
either $a^{\dagger }$,$a$ or $b^{\dagger }$,$b$, 
\begin{equation}
\check{F}_{+}=F_{+}\left( a^{\dagger },a\right) ,\;\check{F}_{-}=F_{-}\left(
b^{\dagger },b\right) .  \label{dm7}
\end{equation}
The mean values of operators $\check{F}_{\pm }$ and the entire information
concerning the subsystems of particles and antiparticles can be obtained
from the so-called reduced density operators, which we shall define below.

We present the basis vectors from (\ref{dm5}) as follows: 
\begin{eqnarray}
&&|\Psi \left( \left\{ m\right\} _{M},\left\{ n\right\} _{N}\right) \rangle
=|\Psi _{a}\left( \left\{ m\right\} _{M}\right) \rangle \otimes |\Psi
_{b}\left( \left\{ n\right\} _{N}\right) \rangle ,\;|0\rangle =|0\rangle
_{a}\otimes |0\rangle _{b}\,,  \notag \\
&&|\Psi _{a}\left( \left\{ m\right\} _{M}\right) \rangle =a_{m_{1}}^{\dagger
}\ldots a_{m_{M}}^{\dagger }|0\rangle _{a}\,,\;|\Psi _{b}\left( \left\{
n\right\} _{N}\right) \rangle =b_{n_{1}}^{\dagger }\ldots b_{n_{N}}^{\dagger
}|0\rangle _{b}\,,  \label{dm9}
\end{eqnarray}
where $|0\rangle _{a}$ and $|0\rangle _{b}$ are the vacuum vectors of
particle and antiparticle subsystems. The mean values of the operators $%
\check{F}_{\pm }$ are 
\begin{equation}
\langle F_{\pm }\rangle =\mathrm{tr}_{+}\mathrm{tr}_{-}\left( \check{F}_{\pm
}\check{\rho}\right) \,,  \label{dm10}
\end{equation}
where $\check{\rho}$ is the density operator of the system, and the reduced
traces $\mathrm{tr}_{\pm }$ of an operator $\check{A}$ are defined as 
\begin{eqnarray}
\mathrm{tr}_{+}\check{A} &=&\sum_{M=0}^{\infty }\sum_{\{m\}}\left( M!\right)
^{-1}\langle \Psi _{a}\left( \left\{ m\right\} _{M}\right) |\check{A}|\Psi
_{a}\left( \left\{ m\right\} _{M}\right) \rangle \,,  \notag \\
\mathrm{tr}_{-}\check{A} &=&\sum_{M=0}^{\infty }\sum_{\{m\}}\left( M!\right)
^{-1}\langle \Psi _{b}\left( \left\{ m\right\} _{M}\right) |\check{A}|\Psi
_{b}\left( \left\{ m\right\} _{M}\right) \rangle \,.  \label{dm11}
\end{eqnarray}
We define the reduced density operators (in the Heisenberg picture) $\check{%
\rho}_{\pm }$ of the subsystems of particles and antiparticles,
respectively, as follows: 
\begin{equation}
\check{\rho}_{\pm }=\mathrm{tr}_{\mp }\check{\rho}\,.  \label{dm13}
\end{equation}
Then mean values (\ref{dm10}) can be calculated with the help of the reduced
density operators $\check{\rho}_{\pm }$ as follows: 
\begin{equation}
\langle F_{\pm }\rangle =\mathrm{tr}_{\pm }\left( \check{F}_{\pm }\check{\rho%
}_{\pm }\,\right) .  \label{dm12}
\end{equation}

Even if the initial state of the system is a pure state, the reduced density
operators $\check{\rho}_{\pm }$ describe mixed states. In some physical
problems, the use of a reduced density operators is inevitable. For example,
considering particle creation by the gravitation field of a black hole we
have only the reduced operator of the particles created outside the black
hole, since we do not have any information about the particles behind the
horizon, \cite{Haw75,FroN89}.

In a similar manner, we introduce the reduced generating operators $\check{R}%
_{\pm }\left( J\right) $ as follows: 
\begin{equation*}
\check{R}_{\pm }\left( J\right) =\mathrm{tr}_{\mp }\check{R}\left( J\right)
\,.
\end{equation*}
Using the path integral representation for traces (\ref{ap12}),
representation (\ref{dm15}), as well as (\ref{dm14c}), we obtain\footnote{%
It should be noted that\ the symbols of the normal form of the operator $%
\hat{R}_{\zeta }$ have been presented in \cite{BuhGF80} via some path
integrals. The explicit form of operator has been presented for $%
J_{n}^{(\zeta )}=e^{-E_{n}^{(\zeta )}}.$ Unfortunately, it contains some
misprints.} 
\begin{eqnarray}
&&\check{R}_{+}\left( J\right) =Z_{+}^{-1}:\exp \left\{ -a^{\dagger }\left(
1-K_{+}\left( J\right) \right) a\right\} :\,,  \notag \\
&&\check{R}_{-}\left( J\right) =Z_{-}^{-1}:\exp \left\{ -b^{\dagger }\left(
1-K_{-}\left( J\right) \right) b\right\} :\,,  \notag \\
&&K_{\pm }\left( J\right) =D_{\pm }+C^{\dagger }\left( 1+\kappa D_{\mp
}^{T}\right) ^{-\kappa }C\,,  \notag \\
&&Z_{\pm }^{-1}\left( J\right) =Z^{-1}\left| c_{v}\right| ^{2}\det \left(
1+\kappa AB\right) ^{\kappa }\det \left( 1+\kappa D_{\mp }\right) ^{\kappa }%
\mathcal{\,}.  \label{dm16}
\end{eqnarray}
The reduced generating operators $\check{R}_{\pm }\left( J\right) $ allow
one to obtain the reduced density operators $\check{\rho}_{\pm }$ for
different initial states of the system. Consider below some examples:

a) Selecting all $J=0$ in (\ref{dm16}), we obtain reduced density operators $%
\check{\rho}_{v\pm }=\check{R}_{\pm }\left( 0\right) $ of a system that has
been in a pure vacuum state at the initial time instant. Explicit
expressions for $\check{R}_{\pm }\left( 0\right) $ follow from (\ref{dm16})
with account taken of 
\begin{equation}
K_{\pm }\left( 0\right) =w\left( +-|0\right) w\left( +-|0\right) ^{\dagger
}\,,\;Z_{\pm }^{-1}\left( 0\right) =\left| c_{v}\right| ^{2}\,.  \label{dm17}
\end{equation}
The same result has been obtained in \cite{BuhGF80,Thermal} by a
straightforward calculation.

b) The reduced density operators $\check{\rho}_{0;n\,\pm }$ and $\check{\rho}%
_{m;0\,\pm }$ of a system that has been in a pure state with particles or
antiparticles, respectively, at the initial time instant can be obtained
from the generating operator$\underline{\check{R}}_{\pm }\left( J\right) =Z%
\check{R}_{\pm }\left( J\right) $ as follows: 
\begin{eqnarray*}
&&\check{\rho}_{m;0\,+}=\left. \frac{\partial \underline{\check{R}}%
_{+}\left( J\right) }{\partial J_{m}^{(+)}}\right| _{J=0}=\left[ a^{\dagger
}w\left( +|+\right) \right] _{m}\check{\rho}_{v+}\left[ w\left( +|+\right)
^{\dagger }a\right] _{m}\;, \\
&&\check{\rho}_{0;m\,-}=\left. \frac{\partial \underline{\check{R}}%
_{-}\left( J\right) }{\partial J_{m}^{(-)}}\right| _{J=0}=\left[ w\left(
-|-\right) b^{\dagger }\right] _{m}\check{\rho}_{v-}\left[ bw\left(
-|-\right) ^{\dagger }\right] _{m}\;, \\
&&\check{\rho}_{0;m\,+}=\left. \frac{\partial \underline{\check{R}}%
_{+}\left( J\right) }{\partial J_{m}^{(-)}}\right| _{J=0}=\check{\rho}_{v+}%
\left[ w\left( -|-\right) w\left( -|-\right) ^{\dagger }\right] _{mm} \\
&&-\left[ a^{\dagger }w\left( +-|0\right) w\left( -|-\right) ^{\dagger }%
\right] _{m}\check{\rho}_{v+}\left[ w\left( -|-\right) w\left( +-|0\right)
^{\dagger }a\right] _{m}\;, \\
&&\check{\rho}_{m;0\,-}=\left. \frac{\partial \underline{\check{R}}%
_{-}\left( J\right) }{\partial J_{m}^{(+)}}\right| _{J=0}=\check{\rho}_{v-}%
\left[ w\left( +|+\right) ^{\dagger }w\left( +|+\right) \right] _{mm} \\
&&-\left[ b^{\dagger }w\left( +-|0\right) w\left( +|+\right) ^{\ast }\right]
_{m}\check{\rho}_{v-}\left[ w\left( +|+\right) ^{T}w\left( +-|0\right)
^{\dagger }b\right] _{m}\;.
\end{eqnarray*}

c) Let us set the sources\emph{\ }in (\ref{dm16})\emph{\ }as in (\ref{dm3}).
One can see that for such a choice of sources the reduced generating
operators (\ref{dm1}) become the reduced density operators $\check{\rho}%
_{\beta \pm }$ of the system that has been in thermal equilibrium at the
initial time instant.

\subsection{One-particle correlation functions}

Let us now examine the following generating functions: 
\begin{eqnarray}
\mathbb{N}_{nm}^{(+)} &=&\mathrm{tr}\left( a_{n}^{\dagger }a_{m}\check{R}%
\right) =\mathrm{tr}_{+}\left( a_{n}^{\dagger }a_{m}\check{R}_{+}\right) \,,
\notag \\
\mathbb{N}_{nm}^{(-)} &=&\mathrm{tr}\left( b_{n}^{\dagger }b_{m}\check{R}%
\right) =\mathrm{tr}_{-}\left( b_{n}^{\dagger }b_{m}\check{R}_{+}\right) \,.
\label{dm20}
\end{eqnarray}
They generate one-particle correlation functions for different initial
states of the system. Setting the sources (taking the corresponding
derivatives, if necessary) in (\ref{dm20}) as has been demonstrated in Sec.
2, we choose the required initial states. The diagonal elements $\mathbb{N}%
_{mm}^{(\zeta )}$ are the generating functionals for the mean numbers $%
N_{m}^{(\zeta )}$ of particles/antiparticles with quantum numbers $m$ at the
final time instant (further differential mean numbers). In what follows, we
refer to quantities (\ref{dm20}) as correlation functions.

The correlation functions $\mathbb{N}_{nm}^{(\zeta )}$ can be expressed via
the matrices $K_{\zeta }$ (\ref{dm16}), and vice-versa, as follows: 
\begin{equation}
\mathbb{N}^{(\zeta )}=\left( \frac{K_{\zeta }}{1+\kappa K_{\zeta }}\right)
^{T}\,,\;K_{\zeta }=\frac{\mathbb{N}^{(\zeta )T}}{1-\kappa \mathbb{N}%
^{(\zeta )T}}\,.  \label{dm23}
\end{equation}
Note that the quantities $K_{\zeta }$ are functions of elementary
probability amplitudes (\ref{a.2a}).

Relations (\ref{dm23}) can be proved as follows: first, using the
commutation relations (\ref{a.1}) we represent (\ref{dm20}) as traces of
operators in the normal form: 
\begin{equation}
\mathbb{N}_{nm}^{(+)}=\mathrm{tr}_{+}\left[ a_{n}^{\dagger }\check{R}%
_{+}\left( K_{+}a\right) _{m}\right] \,,\;\;\mathbb{N}_{nm}^{(-)}=\mathrm{tr}%
_{-}\left[ b_{n}^{\dagger }\check{R}_{-}\left( K_{-}b\right) _{m}\right] \,.
\label{dm22}
\end{equation}
The quantities $\mathbb{N}_{nm}^{(\zeta )}$ can be obtained from the
generating functions $\mathcal{Z}_{\zeta }\left( \bar{j},j\right) $ as
follows: 
\begin{equation}
\mathbb{N}_{nm}^{(\zeta )}=\left. \frac{\partial ^{2}\mathcal{Z}_{\zeta
}\left( \bar{j},j\right) }{\partial \bar{j}_{n}\partial j_{m}}\right| _{\bar{%
j}=j=0}\,,  \label{dm19a}
\end{equation}
where 
\begin{eqnarray}
&&\mathcal{Z}_{+}\left( \bar{j},j\right) =Z_{+}^{-1}\mathrm{tr}_{+}:\exp
\left\{ -a^{\dagger }\left[ 1-\mathbb{I}K_{+}\right] a\right\} :\,,  \notag
\\
&&\mathcal{Z}_{-}\left( \bar{j},j\right) =Z_{-}^{-1}\mathrm{tr}_{-}:\exp
\left\{ -b^{\dagger }\left[ 1-\mathbb{I}K_{-}\right] b\right\} :\,,  \notag
\\
&&\mathcal{Z}_{\zeta }\left( 0,0\right) =1\,,\;\mathbb{I}_{mn}=\delta _{mn}+%
\bar{j}_{m}j_{n}\,,\;\zeta =\pm \,,  \label{dm18}
\end{eqnarray}
and $\bar{j}$ and $j$ are some new sources.{\Large \ }Traces in (\ref{dm18})
can be calculated by formula (\ref{ap12}) from Appendix. Thus, we obtain 
\begin{equation}
\mathcal{Z}_{\zeta }\left( \bar{j},j\right) =Z_{\zeta }^{-1}\exp \left\{
\kappa \sum_{n}\left[ \ln \left( 1+\kappa \mathbb{I}K_{\zeta }\right) \right]
_{nn}\right\} \,.  \label{dm19}
\end{equation}
Then, relations (\ref{dm23}) follow from (\ref{dm19a}) and (\ref{dm19}).

The normalization conditions and the second relation (\ref{dm23}) imply that
the quantities $Z_{\zeta }$ can be expressed in terms of $\mathbb{N}^{(\zeta
)}$ as 
\begin{equation}
Z_{\zeta }=\exp \left\{ \kappa \sum_{n}\left[ \ln \left( 1+\kappa K_{\zeta
}\right) \right] _{nn}\right\} =\exp \left\{ -\kappa \sum_{n}\left[ \ln
\left( 1-\kappa \mathbb{N}^{(\zeta )T}\right) \right] _{nn}\right\} \,.
\label{dm29}
\end{equation}

Now, we are going to relate the quantities $\mathbb{N}_{nm}^{(\zeta )}$ with
the correlation functions $\mathbb{N}_{nm}^{(\zeta )}(in)$ of $in$%
-operators, 
\begin{equation}
\mathbb{N}_{nm}^{(+)}(in)=\mathrm{tr}\left[ a_{n}^{\dagger }(in)a_{m}(in)%
\check{R}\right] \,,\;\mathbb{N}_{nm}^{(-)}(in)=\mathrm{tr}\left[
b_{n}^{\dagger }(in)b_{m}(in)\check{R}\right] \,.  \label{dm24}
\end{equation}
Using representation (\ref{dm1}) for $\check{R}$, one can see that 
\begin{equation}
\mathbb{N}_{nm}^{(\zeta )}(in)=\delta _{nm}N_{m}^{(\zeta
)}(in),\;\;N_{m}^{(\zeta )}(in)=\frac{J_{m}^{(\zeta )}}{1+\kappa
J_{m}^{(\zeta )}}\,,  \label{dm25}
\end{equation}
where $N_{m}^{(\zeta )}(in)$ are the differential mean numbers (generating
functions for differential mean numbers). Indeed, let us take expressions (%
\ref{dm20}) for $\mathbb{N}^{(\zeta )}$ via traces in the complete Fock
space. These traces can be written in the $in$-basis $\left| \Psi \left(
\left\{ m\right\} _{M},\left\{ n\right\} _{N};in\right) \right\rangle
=V\left| \Psi \left( \left\{ m\right\} _{M},\left\{ n\right\} _{N}\right)
\right\rangle $. Using the canonical transformation (\ref{a.9}), we express
the operators $a^{\dagger },a,b^{\dagger },b$ via the operators $a^{\dagger
}(in),a(in),b^{\dagger }(in),b(in)$ and calculate the traces explicitly.
Then we obtain 
\begin{eqnarray}
&&\mathbb{N}^{(+)T}=G\left( ^{+}|_{+}\right) \mathbb{N}^{(+)}(in)G\left(
_{+}|^{+}\right) +G\left( ^{+}|_{-}\right) \left[ 1-\kappa \mathbb{N}%
^{(-)}(in)\right] G\left( _{-}|^{+}\right) ,  \notag \\
&&\mathbb{N}^{(-)}=G\left( ^{-}|_{-}\right) \mathbb{N}^{(-)}(in)G\left(
_{-}|^{-}\right) +G\left( ^{-}|_{+}\right) \left[ 1-\kappa \mathbb{N}%
^{(+)}(in)\right] G\left( _{+}|^{-}\right) .  \label{dm26}
\end{eqnarray}

Thus, due to relations (\ref{dm23}), (\ref{dm26}), we have explicit
expressions for the complete generating density operator (\ref{dm15}) and
reduced generating density operators (\ref{dm16}) via both correlation
functions of $in$-particles and $out$-particles, and via elementary
probability amplitudes (\ref{a.2a}) as well.

We stress that there is a one-to-one correspondence between the one-particle
correlation functions, and the form of the reduced density operator of the
complete system. This correspondence is related to the choice of a model,
which is a quantized field placed in an external background. In fact, we
deal with a quadratic system of noninteracting (between themselves)
particles. Of course, this fact is well-known for free-particle systems. Our
consideration generalizes this desription to the presence of a
particle-creating background. For systems of interacting particles, there
remains an important question: suppose the one-particle distribution at the
final time instant is a thermal one. Can one assert that the complete system
is in a thermal state with a given temperature (one that determines a
one-particle distribution)? Such a question seems to be relevant to the
problem of particle creation by black-hole gravitational fields (Hawking's
radiation), where one-particle distributions of created particles have a
thermal form.

Below, we examine some illustrations of the previously obtained general
formulas.

Let the initial state of the system be vacuum ($J=0$), then (\ref{dm26})
reproduces formulas (\ref{dm27}) for the differential mean numbers $\aleph
_{m}^{(\zeta )}=\left. N_{m}^{(\zeta )}\right| _{J=0}$ of
particles/antiparticles created from vacuum by an external field.

Let us examine a common case (for example, a uniform external field) when
particle/antiparticle states are specified by quantum numbers (the same
being valid for particles and antiparticles) that are integrals of motion.
In this case, all the matrices $G\left( {}_{\zeta }|{}^{\zeta ^{\prime
}}\right) $ in (\ref{emt18}) are diagonal and the differential mean numbers (%
\ref{dm27}) of particles/antiparticles created from vacuum coincide: $\aleph
_{m}^{(+)}=\aleph _{m}^{(-)}=\aleph _{m}$. Using formulas (\ref{dm26}), (\ref%
{dm27}), and the unitarity relations (\ref{vst12}), one can obtain the
expressions for the differential mean numbers of particles/antiparticles,
namely, 
\begin{equation}
N_{m}^{(\zeta )}=\left( 1-\kappa \aleph _{m}\right) N_{m}^{(\zeta
)}(in)+\aleph _{m}\left[ 1-\kappa N_{m}^{(-\zeta )}(in)\right] \,.
\label{dm30}
\end{equation}
If the initial state differs from vacuum, the differential mean numbers of
particles/antiparticles created by the external field are given by the
difference $\Delta N_{m}^{(\zeta )}=N_{m}^{(\zeta )}-N_{m}^{(\zeta )}(in)$.
One can see that 
\begin{eqnarray}
&&\Delta N_{m}^{(+)}=\Delta N_{m}^{(-)}=\Delta N_{m}\,,  \notag \\
&&\Delta N_{m}=\aleph _{m}\left[ 1-\kappa \left(
N_{m}^{(+)}(in)+N_{m}^{(-)}(in)\right) \right] \,.  \label{ex5}
\end{eqnarray}
Even if $\aleph _{m}\neq 0,$ no particle creation of fermions with quantum
numbers $m$ occurs if $N_{m}^{(+)}(in)+N_{m}^{(-)}(in)=1$. Since $\kappa =-1$
for bosons, $\Delta N_{m}$ is always positive and is larger than $\aleph
_{m} $. That is, the presence of matter at the initial state increases the
mean number of created bosons.

In some articles devoted to the chromoelectric flux tube model (see, e.g., 
\cite{NayN05,KhaLT06,GleM84}), one encounters an (inexact) interpretation of
the well-known Schwinger formulas describing pair-creation from vacuum by a
constant electric field \cite{S51}. This interpretation may lead to
incorrect results for some field strengths, as noted in \cite{KluME98}.
Below, we discuss this problem and present correct relations that will be
used in the subsequent section. We recall that, by using the proper-time
method, Schwinger calculated the one-loop effective Lagrangian $L$ in
electric field and assumed that the probability $P^{v}$ of no actual
pair-creation occurring in the history of the field during the time $T$ in
the volume $V$ can be presented as $P^{v}=\left| c_{v}\right| ^{2}=\exp
\{-VT2\func{Im}L\}$ (for a subsequent development, see the review \cite%
{Dunn04}). Schwinger interpreted $2\func{Im}L$ as the probability, per time
unit, and per volume unit, of creating a pair by a constant electric field.
Some arguments in favour of such an interpretation can be found, for
example, in the book \cite{ItzZub80} and the article \cite{CasNN79}. The
interpretation remains approximately valid as long as the WKB calculation is
applicable, that is, $VT2\func{Im}L\ll 1$. Then the total probability of
pair-creation reads as $1-P^{v}\approx VT2\func{Im}L\,.$ To calculate the
differential probabilities of pair-creation with quantum numbers $m$ (for
instance, momentum and spin polarization), one can represent the probability 
$P^{v}$ as an infinite product: 
\begin{equation}
P^{v}=\prod_{m}e^{-2\func{Im}S_{m}}\,,  \label{vst1}
\end{equation}
where a certain discretization scheme is used, so that the effective action $%
S=VTL$ is written as $S=\sum_{m}S_{m}$. All this is possible only if $m$ are
selected as integrals of motion. Then, $e^{-2\func{Im}S_{m}}$ is the
vacuum-persistence probability in a cell of the space of quantum numbers $m$%
. Using the WKB approximation in the case $2\func{Im}S_{m}\ll 1,$ one
obtains for the probability $P_{m}$ of a single pair-production with quantum
numbers $m$ and for the corresponding mean values $\aleph _{m}$ of created
pairs the following relation: 
\begin{equation}
\aleph _{m}\approx P_{m}\approx 2\func{Im}S_{m}\,.  \label{vst2}
\end{equation}

By analogy with one-particle quantum mechanics, one usually rewrites (\ref%
{vst2}) for fermions, 
\begin{equation}
\aleph _{m}\approx -\ln \left( 1-P_{m}\right) \approx 2\func{Im}S_{m}\,.
\label{vst3}
\end{equation}
It is clear that (\ref{vst2}) and (\ref{vst3}) coincide in the first order
with respect to $P_{m}$ . Then, it follows from (\ref{vst1}) that 
\begin{equation}
P^{v}\approx \prod_{m}\left( 1-P_{m}\right) \,.  \label{vst4}
\end{equation}

Using the same analogy for bosons and rewriting (\ref{vst2}) as 
\begin{equation}
\aleph _{m}\approx \ln \left( 1+P_{m}\right) \approx 2\func{Im}S_{m}\,,
\label{vst6}
\end{equation}
one obtains the following approximate relation: 
\begin{equation}
P^{v}\approx \prod_{m}\left( 1+P_{m}\right) ^{-1}\,.  \label{vst5}
\end{equation}

It turns out that for the field under consideration, by using the WKB
calculations and relations (\ref{vst3})--(\ref{vst5}), one can reproduce
Schwinger's result for $P^{v}$. This fact brings the temptation to interpret
the latter formulas as exact ones, replacing there ``$\approx $'' by ``$=$%
''. However, one should say that such an interpretation is, in particular,
equivalent to the assumption $\aleph _{m}=2\func{Im}S_{m}$. Nevertheless, as
we shall demonstrate below, the latter relation is not an exact one and it
is valid only in the approximation $2\func{Im}S_{m}\ll 1$.

An exact treatment in the framework of QFT with unstable vacuum (see, for
example, \cite{GavG96a,Gitman,GavG91}) yields the following expressions for
the scattering $P(-|-)_{m}$ of a particle (and an antiparticle) and
pair-creation $P(+-|0)_{m}$ probabilities, respectively (see Subsection 2.1
for notation): 
\begin{equation}
P(-|-)_{m}=\left| w\left( -|-\right) _{mm}\right|
^{2}P^{v}\,,\;\;P(+-|0)_{m}=\left| w\left( +-|0\right) _{mm}\right|
^{2}P^{v}\,,  \label{vst7}
\end{equation}
where, due to relations (\ref{a.10}), (\ref{dm27}) and (\ref{vst12}), the
corresponding relative probabilities are 
\begin{equation}
\left| w\left( -|-\right) _{mm}\right| ^{2}=\frac{1}{1-\kappa \aleph _{m}}%
\,,\;\;\left| w\left( +-|0\right) _{mm}\right| ^{2}=\frac{\aleph _{m}}{%
1-\kappa \aleph _{m}}\,.  \label{vst8}
\end{equation}
As long as the semiclassical approximation is concerned ($P^{v}\approx 1,$ $%
\aleph _{m}\ll 1$), we have 
\begin{equation*}
P(+-|0)_{m}\approx \left| w\left( +-|0\right) _{mm}\right| ^{2}\approx
\aleph _{m}\,.
\end{equation*}
Thus, we can see that the quantities $P(+-|0)_{m}$, $\left| w\left(
+-|0\right) _{mm}\right| ^{2}$ and $\aleph _{m}$ can be identified only in
the approximation under consideration. An exact expression for $P^{v}$ in
terms of the mean values $\aleph _{m}$ follows from (\ref{a.3c}), (\ref{vst8}%
) and reads 
\begin{equation}
P^{v}=\exp \left\{ \kappa \sum_{m}\ln \left( 1-\kappa \aleph _{m}\right)
\right\} \,.  \label{vst9}
\end{equation}
Formulas (\ref{vst1}) and (\ref{vst9}) imply the following exact relation
between $\func{Im}S_{m}$ and $\aleph _{m}$: 
\begin{equation}
2\func{Im}S_{m}=-\kappa \ln \left( 1-\kappa \aleph _{m}\right) \,.
\label{vst10}
\end{equation}
It has to be used in the general case when the WKB approximation is not
applicable.

\subsection{Is it really a thermal distribution?}

Considerable attention has been recently focused on a mechanism of fast
thermalization in heavy-ion collisions (see \cite{KhaLT06,KhaT05} and
references therein). A possibility is discussed of a thermal one-particle
distribution due to quantum creation of particles from vacuum by strong
electric-like fields. Some of these distributions are known in QED, and
their relation to thermal spectrum of Hawking's radiation has been discussed
(see references in the next section). We give some examples of such
distributions in Subsection 4.3. However, a thermal one-particle
distribution of created particles does not guarantee the character of
thermal equilibrium for the corresponding complete quantum state of the
system and only mimics, in some sense, the latter state.

One ought to say that, in contrast to the case of Hawking's radiation, in
which we do not have any information about another member of each created
pair behind the horizon, both particles and antiparticles created from
vacuum by chromoelectric field can, in principle, be observed. Because of
the one-to-one correspondence between one-particle correlation functions and
the reduced density operator of the complete system, all the moments of
particle (or antiparticle) distribution coincide. However, the higher
moments of the simultaneous distributions of particles and antiparticles are
different.

In what follows, we present a formal analysis of the above problems.

Suppose that the differential mean numbers $N_{m}$ of
particles/antiparticles at the final state of a system subject to an
external field have the form of a one-particle thermal distribution. There
arises the question if one can be sure if in such a case the final state of
the complete system is in thermal equilibrium, or the thermal form of a
one-particle distribution can appear even if the final state of the complete
system is not in thermal equilibrium. To answer these questions, we plan to
examine two different possibilities of having the same one-particle thermal
distribution for two distinct states of the complete system, one of them
being a thermal equilibrium and the other a pure state. Let the first state
of the complete system be described by the thermal density operator: 
\begin{eqnarray}
&&\check{\rho}_{\beta }^{out}=\frac{1}{Z}\exp \left\{ -\left[ a^{\dagger
}E^{(+)}a+b^{\dagger }E^{(-)}b\right] \right\} \,,  \notag \\
&&Z=\exp \left\{ \kappa \sum_{n}\ln \left( 1+\kappa e^{-E_{n}^{(+)}}\right)
+\kappa \sum_{m}\ln \left( 1+\kappa e^{-E_{n}^{(-)}}\right) \right\} \,,
\label{jfm4}
\end{eqnarray}
where $E^{(\pm )}$ are given by (\ref{dm3}). It is obvious that in such a
state the differential mean numbers $N_{m}$ have the form 
\begin{equation}
N_{m}=\left( e^{E_{m}}+\kappa \right) ^{-1}.  \label{xz}
\end{equation}
On the other hand, if we have a causal evolution from vacuum, the density
operator of the corresponding pure state having the form $\check{\rho}_{v}$ (%
\ref{dm1a}); see the normal form in (\ref{dm15a}). Such a state provides the
differential mean numbers (\ref{xz}) in case (\ref{vst8}) holds true. We can
see that measuring the one-particle distribution cannot distinguish between
both these cases. Nevertheless, they can be distinguished by measuring the
next moments, as demonstrated below. Let us calculate the variances $\mathrm{%
Var}_{m}$ in the states described by the density matrices (\ref{jfm4}) and (%
\ref{dm1a}), respectively, 
\begin{eqnarray*}
\mathrm{Var}_{m}^{th} &=&\mathrm{tr}\left[ \left( a_{m}^{\dagger
}a_{m}+b_{m}^{\dagger }b_{m}-2N_{m}\right) ^{2}\check{\rho}_{\beta }^{out}%
\right] , \\
\mathrm{Var}_{m}^{v} &=&\mathrm{tr}\left[ \left( a_{m}^{\dagger
}a_{m}+b_{m}^{\dagger }b_{m}-2N_{m}\right) ^{2}\check{\rho}_{v}\right] .
\end{eqnarray*}
Since the differential mean values coincide in both states, one can see that 
\begin{eqnarray}
&&\mathrm{Var}_{m}^{th}-\mathrm{Var}_{m}^{v}=2\left(
Q_{m}^{th}-Q_{m}^{v}\right) \,,  \notag \\
&&Q_{m}^{th}=\mathrm{tr}\left[ a_{m}^{\dagger }a_{m}b_{m}^{\dagger }b_{m}%
\check{\rho}_{\beta }^{out}\right] \,,\;Q_{m}^{v}=\mathrm{tr}\left[
a_{m}^{\dagger }a_{m}b_{m}^{\dagger }b_{m}\check{\rho}_{v}\right] \,.
\label{jfm5}
\end{eqnarray}
To calculate the quantities $Q_{m}^{th}$, $Q_{m}^{v}$ and demonstrate that
they are actually different, we are going to use the generating functional $%
\Phi ^{v}\left( J\right) $ (\ref{jfm3}), and the generating functional of
momenta for the thermal distribution, 
\begin{equation*}
\Phi ^{th}\left( J\right) =\mathrm{tr}\left\{ \exp \left[ a^{\dagger }%
\mathbb{J}^{(+)}a+b^{\dagger }\mathbb{J}^{(-)}b\right] \check{\rho}_{\beta
}^{out}\right\} .
\end{equation*}
Then 
\begin{equation}
Q_{m}^{th}=\left. \frac{\partial ^{2}\Phi ^{th}}{\partial
J_{m}^{(+)}\partial J_{m}^{(-)}}\right| _{J=0}=N_{m}^{2}\,,  \label{jfm6}
\end{equation}
where the expression 
\begin{equation*}
\Phi ^{th}(J)=\frac{1}{Z}\exp \left\{ \kappa \sum_{n}\ln \left( 1+\kappa
e^{-E_{n}^{(+)}+J_{n}^{(+)}}\right) +\kappa \sum_{m}\ln \left( 1+\kappa
e^{-E_{n}^{(-)}+J_{n}^{(-)}}\right) \right\}
\end{equation*}
is used. On the other hand, 
\begin{equation}
Q_{m}^{v}=\left. \frac{\partial ^{2}\Phi ^{v}}{\partial J_{m}^{(+)}\partial
J_{m}^{(-)}}\right| _{J=0}=N_{m}\left[ 1+\left( 1-\kappa \right) N_{m}\right]
\,.  \label{jfm7}
\end{equation}
Thus, we can see that the variances of the simultaneous distributions of
particles and antiparticles, $\mathrm{Var}_{m}^{th}$ and $\mathrm{Var}%
_{m}^{v}$, are quite different: 
\begin{equation*}
\mathrm{Var}_{m}^{th}-\mathrm{Var}_{m}^{v}=2N_{m}\left( \kappa
N_{m}-1\right) .\,
\end{equation*}

\section{Particle-creation in an electric-like background}

\subsection{Quasi-constant electric field}

Below, we consider a number of applications of the above-developed formalism
in QED with a quasi-constant (slowly varying) uniform electric field
violating the stability of vacuum. We emphasize that our consideration can
be relevant in QCD with an electric-like colour field and in some QFT models
with a curved space-time, as was demonstrated, for example, in \cite{GavGO97}%
. It was shown \cite{GavG96a} that the distribution of pairs created from
vacuum by a quasi-constant electric field has a thermal-like form. It
appears that such a form has a universal character, i.e., it emerges in any
theory with quasi-constant external fields; and when applied to
particle-creation in external constant gravitational fields it reproduces
exactly the Hawking temperature. Thus, our consideration of QED with a
quasi-constant electric field allows one to reveal the typical properties of
any strong-field QFT.

Note that in the case under consideration particle states are specified by
continuous quantum numbers of the momentum $\mathbf{p}$ and spin projections 
$r=\pm 1$ (we formally set $r=0$ for scalar particles). From now on, we
suppose that the standard volume regularization is used, so that $\delta (%
\mathbf{p}-\mathbf{p}^{\prime })$ is replaced by $\delta _{\mathbf{p},%
\mathbf{p}^{\prime }}$ in the normalization conditions. Thus, our particles
are labeled by a set of discrete quantum numbers, $m=(\mathbf{p,}r)$.

As usual, we shall describe the electric field by time-dependent vector
potentials. The states of the quantum system in question are far from
equilibrium due to the field influence. We shall study in detail the time
dependence of various mean values, in particular, the mean values of created
particles. In a physically correct setting of the problem, we consider a
model of a quasi-constant electric field $E(x^{0})$ which effectively acts
only for a finite period of time $T$ and is zero outside this interval (we
further call it the $T$-constant field). In our model, $E(x^{0})=E$ for $%
t_{1}\leq x^{0}\leq t_{2}$, $t_{2}=-t_{1}=T/2$. Thus, the field produces
finite work in a finite space volume. We accept the initial vacuum to be a
free-particle vacuum. A relevant calculation in QED with a $T$-constant
field can be found in \cite{GavG96a}. Below, we use these results for
evaluating the leading terms in particle-creation phenomena at large $T$,
when the effects of switching on and off are negligible.

Let us describe the $T$-constant field. It is nonstationary but has a
constant space direction. We place the latter field along the $x^{3}$-axis.
We denote by $q$ the charge of a particle (by $-q$ that of an antiparticle),
and by $M$ we denote mass. The corresponding potentials can be chosen in the
form: $A_{0}=A_{1}=A_{2}=0,$ and 
\begin{equation*}
A_{3}(x^{0})=\left\{ 
\begin{array}{ll}
Et_{1},\;\; & x^{0}\in I \\ 
Ex^{0},\;\; & x^{0}\in II \\ 
Et_{2},\;\; & x^{0}\in III\;.%
\end{array}
\right.
\end{equation*}
where the time intervals are $I=(-\infty ,t_{1}),$ $II=[t_{1},t_{2}],$ $%
III=(t_{2},+\infty )$.

If the time $T$ is sufficiently large, 
\begin{equation*}
T>>T_{0}=(1+\lambda )/\sqrt{\left| qE\right| }\,,
\end{equation*}
the differential mean numbers $\aleph _{m}$ read 
\begin{equation}
\aleph _{m}=\left\{ 
\begin{array}{l}
e^{-\pi \lambda }\left[ 1+O\left( \left[ \frac{1+\lambda }{K}\right]
^{3}\right) \right] \,,\;\;-\sqrt{\left| qE\right| }\frac{T}{2}\leq \xi \leq
-K\,, \\ 
O\left( 1\right) ,\;\;-K\,<\xi \leq +K\,, \\ 
O\left( \left[ \frac{1+\lambda }{\xi ^{2}}\right] ^{3}\right) ,\;\;\xi >K\,,%
\end{array}
\right.  \label{e40a}
\end{equation}
where $K$ is a sufficiently large arbitrary constant, $K>>1+\lambda $, 
\begin{equation}
\lambda =\frac{M^{2}+\mathbf{p}_{\bot }^{2}}{\left| qE\right| },\;\;\mathbf{p%
}_{\bot }=(p^{1},p^{2},0),\;\;\xi =\frac{\left| p_{3}\right| -\left|
qE\right| T/2}{\sqrt{\left| qE\right| }}\,,  \label{lambda}
\end{equation}
and $p_{3}$ is a longitudinal momentum of a particle \cite{GavG96a}. One can
examine the limit $T\rightarrow \infty $ at any given $\mathbf{p}$ in the
above expression. In such a limit, the differential mean numbers have a
simple form: 
\begin{equation}
\aleph _{m}=e^{-\pi \lambda }  \label{e40ad}
\end{equation}
which coincides with that obtained in a constant electric field by Nikishov 
\cite{Nikis79}. One can see that the stabilization of the differential mean
numbers to the asymptotic form (\ref{e40ad}) for finite longitudinal momenta
is reached at $T>>T_{0}$. The characteristic time $T_{0}$ is called the
stabilization time.

In order to investigate the effects of switching on and off for $T>>T_{0}$,
we are going to consider a different example of a quasi-constant electric
field: 
\begin{equation}
E(x^{0})=E\cosh ^{-2}\left( \frac{x^{0}}{\alpha }\right) \,.  \label{e2a}
\end{equation}
This field switches on and off adiabatically as $x^{0}\rightarrow \pm \infty 
$ and is quasi-constant at finite times. It is called an adiabatic field.
The differential mean numbers of particles created by such a field have been
found in \cite{NarN70}. For a further discussion, we need these numbers for
a large $\alpha .$ As has been demonstrated in \cite{GavG96a}, the
differential mean numbers in the field (\ref{e2a}) take the asymptotic form (%
\ref{e40ad}) for $\alpha >>\alpha _{0}=(1+\sqrt{\lambda })/\sqrt{\left|
qE\right| }$ and for $|p_{3}|<<\left| qE\right| \alpha $. Thus, $\alpha _{0}$
can be interpreted as the stabilization time for an adiabatic field. At the
same time, the latter fact means that the effects of switching on and off
are not essential at large times and finite longitudinal momenta for both
fields. Extrapolating this conclusion, one can suppose that
particle-creation effects in any electric field, that is one being
quasi-constant $\approx E$ at least for a time period $T>>T_{0}$ and
switching on and off outside this period arbitrarily, do not depend on the
details of switching on and off. Thus, our calculations in a $T$-constant
field are typical for a large class of quasi-constant electric fields.

It is of interest for phenomenological applications to calculate the
distribution of particles created with all possible $p_{3}$ values and a
given $p_{\perp }$ (it is called the $p_{\perp }$ distribution and is
denoted by $n_{p_{\perp }}$, in what follows). Analysing the total mean
number of particles created by the $T$-constant field, we go over from
summation to integration: $\sum_{\mathbf{p}}\rightarrow \frac{V}{\left( 2\pi
\right) ^{3}}\int d\mathbf{p}$. Then, the total mean number (we denote it by 
$\aleph $) can be presented as 
\begin{equation}
\aleph =V\int d^{2}p_{\perp }n_{p_{\perp }},  \label{4.1.1}
\end{equation}
where 
\begin{equation}
n_{p_{\perp }}=\frac{1}{(2\pi )^{3}}\sum_{r}\int dp_{3}\aleph _{m}
\label{4.1.2}
\end{equation}
is the $p_{\perp }$ distribution density of particles created per unit
volume. $\aleph _{m}$ is constant for $\left| p_{3}\right| \leq \sqrt{\left|
qE\right| }\left( \sqrt{\left| qE\right| }T/2-K\right) $ and for $T>>T_{0}$,
and decreases rapidly for $\left| p_{3}\right| >\sqrt{\left| qE\right| }%
\left( \sqrt{\left| qE\right| }T/2+K\right) $. The contribution to the
integral (\ref{4.1.2}) from the intermediate region can be estimated as $2%
\sqrt{\left| qE\right| }K$. This implies 
\begin{equation}
n_{p_{\perp }}=\frac{J\sqrt{\left| qE\right| }}{(2\pi )^{3}}\left[ \sqrt{%
\left| qE\right| }Te^{-\pi \lambda }+O(K)\right] ,  \label{4.1.3}
\end{equation}
where $J$ is the number of the spin degrees of freedom ($J=1$ for scalar
particles and $J=2$ for fermions). Thus, the $p_{\perp }$ distribution
density of the particle production rate has the form 
\begin{equation}
\frac{dn_{p_{\perp }}}{dT}=\frac{J\left| qE\right| }{(2\pi )^{3}}e^{-\pi
\lambda }\,.  \label{4.1.4}
\end{equation}
The total number of particles created per unit volume is given by 
\begin{equation}
\frac{\aleph }{V}=J\frac{\left( qE\right) ^{2}T}{\left( 2\pi \right) ^{3}}%
\exp \left\{ -\pi \frac{M^{2}}{\left| qE\right| }\right\} .  \label{4.1.5}
\end{equation}

Suppose that, in addition to the external electric field, there exists a
parallel constant magnetic field $B.$ For definiteness, let us choose its
potentials as $A_{\mu }^{B}=Bx^{2}\delta _{\mu }^{1}$. Then, the complete
set of quantum numbers that describes particles in such a background is $%
(p_{1},n_{B},p_{3},r)$, $n_{B}=0,1,\ldots $, and $\lambda =\left(
M^{2}+|qB|(2n_{B}+1-r)\right) \left| qE\right| ^{-1}$. Substituting this $%
\lambda $ into eqs. (\ref{e40ad}) and (\ref{4.1.3}), we obtain the
differential mean numbers and $p_{\perp }$-distribution density of particles
created per unit volume, respectively. One can see that the presence of the
magnetic field essentially changes the energy spectrum of the transversal
modes (which used to be $\mathbf{p}_{\bot }^{2}$ in the absence of the
magnetic field) and also the form of $\lambda $, but does not change the
dependence of $\aleph _{m}$ and $n_{p_{\perp }}$on $\lambda $. This is
related to the fact that the magnetic field itself does not produce work
acting on charged particles, and, therefore, does not create particles.
Then, the total number of particles created per unit volume is given by 
\begin{equation}
\frac{\aleph }{V}=\frac{Jq^{2}EBT[\cosh (\pi B/E)]^{(1+\kappa )/2}}{8\pi
^{2}\sinh (\pi B/E)}\exp \left\{ -\pi \frac{M^{2}}{\left| qE\right| }%
\right\} \smallskip ,  \label{emt45}
\end{equation}
where $\kappa =+1$ for fermions and $\kappa =-1$ for scalar particles.

For a strong electric field, $M^{2}/\left| qE\right| \lesssim 1$, ($B=0$),
and large $T,$ the energy density of created pairs reads $\mathcal{E}=\left|
qE\right| T\aleph /V$; see \cite{Gav06}. We can neglect the back-reaction of
particles created by the electric field in case their energy density is
essentially smaller than the energy density of the electric field, $\mathcal{%
E}\ll E^{2}/8\pi $. Consequently, the concept of a strong constant electric
field is consistent only if the following condition holds true: 
\begin{equation}
1\ll \left| qE\right| T^{2}\ll \frac{\pi ^{2}}{Jq^{2}}\exp \left\{ \pi \frac{%
M^{2}}{\left| qE\right| }\right\} .  \label{val}
\end{equation}

Following \cite{GavG96a}, we represent the asymptotic formula (\ref{e40ad})
in a universal form: 
\begin{equation}
\aleph _{m}=\exp \left\{ -2\pi \frac{\omega _{m}}{g}\right\} \,,  \label{ct1}
\end{equation}
where $\omega _{m}$ is the work of an external field creating a particle
from a pair in a given state $m$, 
\begin{equation*}
\omega _{m}=\frac{1}{2}\left[ p_{0}(t_{f})+p_{0}(t_{i})+\Delta \epsilon
_{vac}\right] \,,
\end{equation*}
where $p_{0}(t_{f})$ and $p_{0}(t_{i})$ are particle energies at the final
time instant $t_{f}$ and at an initial time instant $t_{i}$, respectively,
and $\Delta \epsilon _{vac}$ is a shift of the vacuum energy due to the time
evolution. The quantity $g$ is the classical acceleration of a particle at
the final time instant. In the case of a $T$-constant field, one can obtain 
\begin{equation*}
\omega _{m}=\frac{M^{2}+\mathbf{p}_{\bot }^{2}}{2p_{0}(t_{f})}=\frac{\lambda 
}{T}\,,\;\;g=\frac{\left| qE\right| }{p_{0}(t_{f})}=\frac{2}{T}\,.
\end{equation*}
Thus, we can see that the differential mean values (\ref{ct1}) are given, in
fact, by the Boltzmann formula with the temperature $\theta =\frac{g}{2\pi
k_{B}}$ (where $k_{B}$ is the Boltzmann constant), the latter having
literally the Hawking form \cite{Haw75}; see below.

We recall that the Hawking result for bosons created by the static
gravitational field of a black hole in a specific thermal environment has
the Planck form 
\begin{equation}
\aleph _{m}=\left[ \exp \left\{ 2\pi \frac{\omega _{m}}{g_{(H)}}\right\} -1%
\right] ^{-1}\,.  \label{ec1}
\end{equation}
Here, $\omega _{m}$ is the energy of a created particle and the Hawking
temperature reads $\theta _{(H)}=\frac{g_{(H)}}{2\pi k_{B}}$, where $g_{(H)}=%
\frac{GM}{r_{g}^{2}}$ is the free-fall acceleration at the gravitational
radius $r_{g}$ of a black hole with mass $M$. In this case of a quasi-static
gravitation field, the evolution shift of the vacuum energy is $\Delta
\epsilon _{vac}=0$, so that one identifies the work $\omega _{m}$ (we have
introduced) with the energy of a particle in formula (\ref{ec1}). It is also
known \cite{Rindler} that an observer which moves with a constant
acceleration $g_{(R)}$ (with respect to its proper time) will probably
register in the Minkowski vacuum some particles (Rindler particles). The
distribution of Rindler bosons has the same Planck form (\ref{ec1}), where
one needws to replace $g_{(H)}$ by $g_{(R)}$, so that the corresponding
temperature is $\theta _{(R)}=\frac{g_{(R)}}{2\pi k_{B}}$.

It is a direct consequence of the equivalence principle that the effective
temperature $\theta $ of distribution (\ref{ct1}) has literally the Hawking
form. The different form of distributions can be caused by essentially
different structures of the Fock space in both cases. We believe that the
Planck distribution arises necessarily due to the appearance of an event
horizon (there is a boundary of the domain of the Hamiltonian), that is, due
to the condition for which the space domains of the particle and
antiparticle vacua are not the same. On the other hand, the final state can
be treated as an equilibrium state. In contrast to this, in a uniform
electric field we deal, in fact, with both the particle vacuum and the
antiparticle vacuum defined over the entire space, that is, these space
domains coincide. In this case, the mixed state of particles (antiparticles)
described by the $\check{\rho}_{v+}$ ($\check{\rho}_{v-}$) density matrix of
Sec. 3 can be represented as a pure state in an extended phase space where
the space domains of both the particle vacuum and the antiparticle vacuum
are the same, being a state of a far-from-equilibrium system. Let us note
that in the framework of a semi-classical description at $\omega _{m}/g<<1$
the Boltzmann spectrum closely approaches the Planck spectrum.

\subsection{ Soft parton production by SU(3) chromoelectric field}

As mentioned in Introduction, in QCD there exist physical situations that
are quite efficiently described by the chromoelectric flux tube model. In
this model, the back-reaction of created pairs induces a gluon mean field
and plasma oscillations (see \cite{KluES93} and references therein). It
appears that the calculation of particle-creation in this model requires the
application of the general formalism of QFT for pair-production at a finite
temperature and at zero temperature both from vacuum and from many-particle
states (see \cite{GanKP95,KluME98,Sch98} for physical reasons). The study of
various time scales in heavy-ion collisions shows that the stabilization
time $T_{0}$ is far smaller than the period of plasma and mean-field
oscillations. Then, according to condition (\ref{val}), the approximation of
a strong $T$-constant chromoelectric field can be used in treating such
collisions during a period when the produced partons can be considered as
weakly coupled due to the property of asymptotic freedom in QCD. It may also
be reasonable to neglect dynamical back-reaction effects and to consider
only pair-production from vacuum by a constant $SU(3)$ chromoelectric field.

Here, we would like to turn our attention to results obtained in QCD with a
constant $SU(3)$ chromoelectric field $E^{a}$ ($a=1,\ldots ,8$) along the $%
x^{3}$-axis; see\ \cite{NayN05}. In this work, the imaginary parts of
one-loop effective actions for quarks $S^{quark}$ and gluons $S^{gluon}$
have been calculated via gauge-invariant $p_{\perp }$ distributions $%
S_{p_{\perp }}^{quark}$ and $S_{p_{\perp }}^{gluon}$ respectively. They have
the form 
\begin{eqnarray}
&&\func{Im}S^{quark}=\int d^{2}p_{\perp }\func{Im}S_{p_{\perp
}}^{quark}\,,\;\;\func{Im}S^{gluon}=\int d^{2}p_{\perp }\func{Im}S_{p_{\perp
}}^{gluon}\,,  \notag \\
&&\func{Im}S_{p_{\perp }}^{quark}=-\frac{VT}{8\pi ^{3}}\sum_{j=1}^{3}\left|
qE_{(j)}\right| \ln \left( 1-e^{-\pi \lambda _{(j)}}\right) \,,  \notag \\
&&\func{Im}S_{p_{\perp }}^{gluon}=\frac{VT}{8\pi ^{3}}\sum_{j=1}^{3}\left| q%
\tilde{E}_{(j)}\right| \ln \left( 1+e^{-\pi \tilde{\lambda}_{(j)}}\right) \,,
\notag \\
&&\lambda _{(j)}=\frac{M^{2}+\mathbf{p}_{\bot }^{2}}{\left| qE_{(j)}\right| }%
,\;\;\tilde{\lambda}_{(j)}=\frac{\mathbf{p}_{\bot }^{2}}{\left| q\tilde{E}%
_{(j)}\right| }\,,  \label{su1}
\end{eqnarray}
where $E_{(j)}$ are the eigenvalues of the matrix $iT^{a}E^{a}$ for the
fundamental representation of $SU(3)$; $\tilde{E}_{(j)}$ are the positive
eigenvalues of the matrix $if^{abc}E^{c}$ for the adjoint representation of $%
SU(3)$; and $q$ is the coupling constant. These eigenvalues are the
following gauge-invariant quantities: 
\begin{eqnarray*}
E_{(1)} &=&\sqrt{C_{1}/3}\cos \theta \,,\;E_{(2)}=\sqrt{C_{1}/3}\cos \left(
2\pi /3-\theta \right) \,, \\
E_{(3)} &=&\sqrt{C_{1}/3}\cos \left( 2\pi /3+\theta \right) \,,
\end{eqnarray*}
where $\theta $ is given by $\cos ^{2}3\theta =3C_{2}/C_{1}^{3},$ and 
\begin{eqnarray*}
\tilde{E}_{(1)} &=&\left[ \frac{C_{1}}{2}\left( 1-\cos \tilde{\theta}\right) %
\right] ^{1/2},\;\;\tilde{E}_{(2)}=\left[ \frac{C_{1}}{2}\left( 1+\cos
\left( \frac{\pi }{3}-\tilde{\theta}\right) \right) \right] ^{1/2}, \\
\tilde{E}_{(3)} &=&\left[ \frac{C_{1}}{2}\left( 1+\cos \left( \frac{\pi }{3}+%
\tilde{\theta}\right) \right) \right] ^{1/2},
\end{eqnarray*}
where $\tilde{\theta}$ is given by $\cos ^{3}\tilde{\theta}%
=-1+6C_{2}/C_{1}^{3}$. Here, $C_{1}$ and $C_{2}$ are Casimir invariants for $%
SU(3)$, 
\begin{equation*}
C_{1}=E^{a}E^{a},\;\;C_{2}=\left( d_{abc}E^{a}E^{b}E^{c}\right) ^{2},
\end{equation*}
where $d_{abc}$ is a symmetric invariant tensor in the adjoint
representation of $SU(3)$. Then, the probabilities $P^{v}$ for a vacuum to
remain a vacuum are found, for both quarks and gluons, from relation (\ref%
{vst1}). However, formulas for parton production rates obtained in \cite%
{NayN05} hold only in the approximation $2\func{Im}S_{p_{\perp }}^{quark}\ll
1$ and $2\func{Im}S_{p_{\perp }}^{gluon}\ll 1$ by virtue of the arguments
that we present at the end of Sbsection 3.2. To obtain exact results, we can
use the following line of reasoning. The results\ (\ref{su1}) can be treated
as those obtained in the case of a $T$-constant chromoelectric field when
the integration over the longitudinal momentum and the summation over the
spin and color degrees of freedom have been carried out. Then, using
relation (\ref{vst10}), we can extract from representation (\ref{su1}) an
exact expression for $p_{\perp }$ distribution densities of quarks $%
n_{p_{\perp }}^{quark}$ and gluons $n_{p_{\perp }}^{gluon}$ produced per
unit volume. Those are 
\begin{equation}
n_{p_{\perp }}^{quark}=\frac{T}{4\pi ^{3}}\sum_{j=1}^{3}\left|
qE_{(j)}\right| e^{-\pi \lambda _{(j)}}\,,\;\;n_{p_{\perp }}^{gluon}=\frac{T%
}{4\pi ^{3}}\sum_{j=1}^{3}\left| q\tilde{E}_{(j)}\right| e^{-\pi \tilde{%
\lambda}_{(j)}}\,,  \label{su2}
\end{equation}
where $T$ is a sufficiently large action period of a constant field. The $%
p_{\perp }$ distribution densities of particle production rates can be found
as $dn_{p_{\perp }}^{quark}/dT$ and $dn_{p_{\perp }}^{gluon}/dT$,
respectively. The total numbers of quarks and gluons created per unit volume
can be obtained from (\ref{su2}) as follows: 
\begin{equation}
\frac{\aleph ^{quark}}{V}=\frac{T}{4\pi ^{3}}\sum_{j=1}^{3}\left(
qE_{(j)}\right) ^{2}\exp \left\{ -\pi \frac{M^{2}}{\left| qE_{(j)}\right| }%
\right\} ,\;\;\frac{\aleph ^{gluon}}{V}=\frac{3Tq^{2}C_{1}}{8\pi ^{3}}\,,
\label{su3}
\end{equation}
where the relation $\sum_{j=1}^{3}\tilde{E}_{(j)}^{2}=3C_{1}/2$ is used.
Taking into account the relation $\sum_{j=1}^{3}E_{(j)}^{2}=C_{1}/2,$ one
can see from (\ref{su3})\ that in a sufficiently strong field $E_{(j)}$, $%
M^{2}/\left| qE_{(j)}\right| \ll 1$, the densities of created quarks and
gluons are related by $\aleph ^{quark}/V=\aleph ^{gluon}/3V$.

We can see that the $(j)$- terms in expressions (\ref{su1}) and (\ref{su3})
can be interpreted as those which are obtained for Abelian-like electric
fields $E_{(j)}$ and $\tilde{E}_{(j)}$, respectively. The maxima of the
fields are restricted by the conditions $\left| E_{(j)}\right| \leq \sqrt{%
C_{1}/3\text{ }}$ and $\left| \tilde{E}_{(j)}\right| \leq \sqrt{C_{1}}$.
Therefore, in order to study the validity of the constant $SU(3)$
chromoelectric field approximation we need to take into account only the
energy density of gluons created by the field $\tilde{E}_{(j)}$. We know
from the previous subsection that this energy per a single pair is $\left| q%
\tilde{E}_{(j)}\right| T$. Then, the total energy density of created gluons
reads 
\begin{equation*}
\mathcal{E}=\frac{T^{2}}{4\pi ^{3}}\sum_{j=1}^{3}\left| q\tilde{E}%
_{(j)}\right| ^{3}\lesssim \left| q\right| \sqrt{C_{1}}T\frac{\aleph ^{gluon}%
}{V}\,.
\end{equation*}
One can neglect the back-reaction of these gluons created by the
chromoelectric field only if $\mathcal{E\ll }C_{1}/8\pi $. Finally, the
condition of validity of the $T$-constant $SU(3)$ chromoelectric field
approximation can be written as 
\begin{equation}
1\ll \left| q\right| \sqrt{C_{1}}T^{2}\ll \frac{\pi ^{2}}{3q^{2}}\,.
\label{su4}
\end{equation}

Therefore, we can see that the $T$-constant $SU(3)$ chromoelectric field
approximation is consistent during the period when the produced partons can
be treated as weakly coupled.

Recently, it has been discovered (see, \cite{LapML06}), in the framework of
the Color Glass Condensate Approach to the description of heavy ion
collisions, that shortly after the collision the system contains both
longitudinal chromoelectric and chromomagnetic fields. In such a
chromomagnetic field, the above-described soft parton production is changed.
First of all, the energy spectrum of the transversal modes becomes
different, which implies that the expressions for $\mathbf{p}_{\bot }^{2}$
that enter $\lambda _{(j)}$ and $\tilde{\lambda}_{(j)}$ are different from
the case of pure chromoelectric field. For example, if the chromomagnetic
field $B^{c}$ is uniform then $\mathbf{p}_{\bot }^{2}=|qB^{c}|(2n_{B}+1-r)$, 
$n_{B}=0,1,\ldots $. This expression is similar to the case of QED, see
Section 4.1. Here, the spin quantum number $r$ takes the values $\pm 1$ for
quarks and $-2,0,+2$ for gluons. As in QED, the constant chromomagnetic
field does not produce work acting on charged particles, and, therefore,%
\textrm{\ }does not create particles. The distribution densities of created
quarks $n_{p_{\perp }}^{quark}$ and gluons $n_{p_{\perp }}^{gluon}$ as
functions of $\lambda _{(j)}$ and $\tilde{\lambda}_{(j)}$ do not change and
are given by equations (\ref{su2}).

Nevertheless, the above expression for $\mathbf{p}_{\bot }^{2}$ in a uniform
chromomagnetic field is negative for $n_{B}=0,$ $r=+2$. This is why such a
configuration of the chromomagnetic field is unstable under quantum
fluctuations. For the first time,\ it was mentioned in \cite{NieO78}. Then,
the problem has been discussed in numerous articles: for a review, see,
e.g., \cite{AmbO80,Flo83}. One of the possible stable configurations of the
longitudinal chromomagnetic field is the so-called flux-tube state, alias
the spaghetti state (see \cite{AmbO80} and references therein). An explicit
form of the spectra $\mathbf{p}_{\bot }^{2}$ in this field is unknown.
Nevertheless, a stable chromomagnetic field is constant and longitudinal.
This is sufficient for us to make the above conclusion that the dependence
on $\lambda _{(j)}$ and $\tilde{\lambda}_{(j)}$ , given by eq. (\ref{su2}),
in the presence of a longitudinal chromomagnetic field remains unchanged.

In the following sections, we turn once again to particle creation by
electric field in QED. The above discussion shows that it can be useful for
understanding the effects of quark and gluon creation in QCD.

\subsection{Thermal-like distributions}

As has been mentioned, the thermalization stage of multiparticle production
in ion-ion collisions at high energies is very important. On the other hand,
as we know from Section 3, it is sometimes difficult to distinguish a real
thermal equilibrium from a state where we have a one-particle thermal
distribution. In this connection, we shall consider some simple examples
when pair-creation by electric field can mimic a one-particle thermal
distribution.

We recall that due to the screen of created pairs, the original electric
field may have an exponential fall-off: 
\begin{equation}
E(x^{0})=Ee^{-x^{0}/\alpha }\,.  \label{tld1}
\end{equation}
The differential mean number of particles created from vacuum by this field
has been calculated in \cite{Spo82}. The result is 
\begin{equation}
\aleph _{m}=\left\{ 
\begin{array}{l}
\left[ \frac{\cosh \left[ \pi \alpha \left( \varepsilon +p_{3}^{\prime
}\right) \right] }{\cosh \left[ \pi \alpha \left( \varepsilon -p_{3}^{\prime
}\right) \right] }e^{2\pi \alpha \varepsilon }-1\right] ^{-1}\;\;\mathrm{Bose%
}\;\mathrm{case,} \\ 
\left[ \frac{\sinh \left[ \pi \alpha \left( \varepsilon +p_{3}^{\prime
}\right) \right] }{\sinh \left[ \pi \alpha \left( \varepsilon -p_{3}^{\prime
}\right) \right] }e^{2\pi \alpha \varepsilon }+1\right] ^{-1}\;\;\mathrm{%
Fermi\;case,}%
\end{array}
\right.  \label{tld2}
\end{equation}
where $\varepsilon =\sqrt{M^{2}+\mathbf{p}^{2}}$ and $p_{3}^{\prime }=p_{3}%
\mathrm{sgn}(qE)$. For $\pi \alpha \left| p_{3}\right| \ll 1$, these
expressions coincide with the Bose and Fermi distributions at the
temperature $\theta =\left( 2\pi k_{B}\alpha \right) ^{-1}$, respectively.

Another example is the pulse of electric field (\ref{e2a}). The differential
mean number of particles created from vacuum by the sharp field pulse (\ref%
{e2a}) at $\alpha \left| qE\right| /\varepsilon \ll 1$ can be extracted from
the result \cite{NarN70} and has the form 
\begin{equation}
\aleph _{m}=\left\{ 
\begin{array}{l}
\left( \pi qE\alpha ^{2}\right) ^{2}\left[ \left( qE\alpha ^{2}\right)
^{2}+\left( p_{3}/\varepsilon \right) ^{2}\right] \sinh ^{-2}\left( \pi
\alpha \varepsilon \right) \;\;\mathrm{Bose\;case,} \\ 
\left( \pi qE\alpha ^{2}\right) ^{2}\left[ 1-\left( p_{3}/\varepsilon
\right) ^{2}\right] \sinh ^{-2}\left( \pi \alpha \varepsilon \right) \;\;%
\mathrm{Fermi\;case,}%
\end{array}
\right.  \label{tld3}
\end{equation}
When the ratio $\left| p_{3}\right| /\varepsilon $ is sufficiently small and
the external field is not strong, $\varepsilon \left/ \sqrt{\left| qE\right| 
}\gg 1\right. $, there is a range of values $\alpha $, $\pi \alpha \epsilon
\gg 1$, in which distributions (\ref{tld3}) have the Boltzmann form with the
temperature $\theta =\left( 2\pi k_{B}\alpha \right) ^{-1}$.

As we note in the previous subsection, depending on the details of the
chromoelectric flux tube model, its stage and field strength, pair
production both from vacuum and from many-particle states by a $T$-constant
electric field may be relevant. The well-known asymptotic form (\ref{e40ad})
of the differential mean numbers for pairs created from vacuum by a constant
electric field can lead to a thermal-like distribution of created pairs if,
for example, the chromoelectric string tension undergoes Gaussian
fluctuations \cite{Bia99}. This implies a modification of the original flux
tube model by introducing a fluctuating string tension. In case the original
flux tube model still holds, nevertheless, the differential mean numbers of
pairs created by\ a $T$-constant electric field can be represented as the
Boltzmann distribution (\ref{ct1}) with the temperature $\theta =\left( \pi
k_{B}T\right) ^{-1}$, as has been seen in Subsection 4.1. Then, it may be
reasonable to examine the phenomenological model with a slowly oscillated
mean electric field and suppose that the pair creation by the mean field
during the semiperiod of oscillation can be effectively approximated by a $T$%
-constant electric field (we recall that the time scale of stabilization $%
T_{0}$ is far smaller than the period of oscillations). In this case, the
electric field produces pairs for a semi-period of oscillation in the
presence of pairs created at previous stages. This is the way to take into
account the effects of back-reaction in such a model. In other words, we are
going to consider pair-creation from the initial state given by a
distribution of previously created particles. Formula (\ref{ex5}) is
relevant in this analysis.

Starting from the initial vacuum state, one has $\Delta N_{m}=\aleph _{m}$,
where $\aleph _{m}$ belongs to the asymptotic form (\ref{e40ad}). Then, at
the end of the first stage, when the mean field is depleted for the first
time, the distribution of particles (being equal to that of antiparticles)
is $N_{m}^{(1)}=\aleph _{m}\,.$ During the second stage, the direction of
the mean field is opposite to the field direction at the first stage. Due to
the condition of stabilization, this is of no importance, since $\aleph _{m}$
is an even function of $qE$. Thus, when the mean field is depleted for the
second time, from (\ref{ex5}) there follows the equality 
\begin{equation*}
N_{m}^{(2)}=\aleph _{m}+\left( 1-2\kappa \aleph _{m}\right) N_{m}^{(1)}\,,
\end{equation*}
and at the end of the $n$-th stage, 
\begin{equation*}
N_{m}^{(n)}=\aleph _{m}+\left( 1-2\kappa \aleph _{m}\right) N_{m}^{(n-1)}\,.
\end{equation*}
Consequently, the total number of particles created at the end of the $n$-th
stage is 
\begin{equation*}
N_{m}^{(n)}=\aleph _{m}\sum_{l=0}^{n-1}\left( 1-2\kappa \aleph _{m}\right)
^{l}\,.
\end{equation*}
We have this result if the created particles do not leave the region of the
active field. In order to take into account the possible loss of particles
due to interaction, movement, etc., we also assume that the total number of
particles at the initial state of the $n$-th stage is less than the number $%
N^{(n-1)}$ of particles created at the end of the $\left( n-1\right) $-th
stage and is $\gamma N^{(n-1)}$, where $\gamma <1$ is the factor of loss.
Then, the modified relation is 
\begin{equation}
N_{m}^{(n)}=\aleph _{m}+\left( 1-2\kappa \aleph _{m}\right) \gamma
N_{m}^{(n-1)}\,,  \label{mpc1}
\end{equation}
and we finally have 
\begin{equation}
N_{m}^{(n)}=\aleph _{m}\sum_{l=0}^{n-1}\gamma ^{l}\left( 1-2\kappa \aleph
_{m}\right) ^{l}\,.  \label{mpc2}
\end{equation}
Supposing that $\gamma $ is constant, one can calculate the sum in (\ref%
{mpc2}), 
\begin{equation}
N_{m}^{(n)}=\aleph _{m}\frac{1-r^{n}}{1-r},\;r=\gamma \left( 1-2\kappa
\aleph _{m}\right) \,.  \label{mpc3}
\end{equation}

For fermions, $\kappa =+1$, then $N_{m}^{(n)}\leq 1$. Energy dissipation
after a period of oscillation is estimated (for real parameters of heavy-ion
collisions) not to be large, so that damping is small and the number of
oscillations can be quite large; damping decreases with an increasing field
strength. If the number of cycles is sufficiently large, we get the limiting
thermal-like distribution 
\begin{equation}
N_{m}^{\Sigma }=\frac{\aleph _{m}}{1-\gamma \left( 1-2\kappa \aleph
_{m}\right) }=\frac{1}{2\gamma }.\frac{1}{e^{\pi \lambda }\left( 1-\gamma
\right) /2\gamma +\kappa }\,.  \label{mpc4}
\end{equation}
In other words, the system reaches a quasi-equilibrium state. For bosons, $%
\kappa =-1$, then $N_{m}^{(n)}$ increases. This is the phenomenon of
resonance, and the increase can be either limited or unlimited depending on
the factor $\gamma $. The increase is limited as long as $r<1$. In this
case, formula (\ref{mpc4}) is valid for bosons as well. We can see that
back-reaction-induced plasma oscillations can reach a quasi-stationary form
specified by thermal-like distribution for both bosons and fermions.

\subsection{Particle creation at finite temperature}

We are now ready to present explicitly the mean number of (anti)particles in
the mode $m$ (with finite longitudinal momenta, $\left| p_{3}\right| \leq 
\sqrt{\left| qE\right| }\left( \sqrt{\left| qE\right| }T/2-K\right) $ ) for
the final state of evolution in a quasi-constant field from the initial
thermodynamical equilibrium, $N_{m}^{(\zeta )}(in)=\left( e^{E_{m}}+\kappa
\right) ^{-1}$, at equal chemical potentials $\mu ^{(+)}=\mu ^{(-)}=\mu $, ($%
\mu <M$ for bosons), 
\begin{equation}
N_{m}^{(\zeta )}=\left( e^{E_{m}}+\kappa \right) ^{-1}+e^{-\pi \lambda
}\left( \tanh \left( E_{m}/2\right) \right) ^{\kappa }\,,  \label{dm31}
\end{equation}
where $E_{m}=\beta \left( \varepsilon _{m}-\mu \right) $, $\varepsilon _{m}=%
\sqrt{M^{2}+\mathbf{p}_{\perp }^{2}+\left( \pi _{3}\right) ^{2}}$, $\pi
_{3}=p_{3}+qET/2$, and it is implied that $\lambda $ is given by (\ref%
{lambda}). This result for the electric field coincides with the one
obtained in \cite{BuhGF80}. Due to the effect of stabilization, it seems
that the time-dependence of the final distributions in question is absent.
However, the integral mean numbers vary as long as a quasi-constant field is
active.

It is of interest to establish a general behavior of the integral mean
numbers of created particles when the effects of switching on and off are
negligible. As has been shown above, we can satisfy this condition by
selecting the action time $T$ of the $T$-constant field ($T>>T_{0}$) as an
effective period of pair creation. It is implied that, in general, the final
time instant, $t_{f}$, and the initial time instant, $t_{i}$, are so
selected that the quasi-constant field is closely approximated by the $T$%
-constant field for a period from $t_{i}$ to $t_{f}$, and $t_{f}-t_{i}=T$.

Let us estimate the sum over the longitudinal momentum $p_{3}$ of $\Delta
N_{m}$ in (\ref{ex5}), which is the mean number of particles created with
all the possible values $p_{3}$. As above, $\sum_{\mathbf{p}}\rightarrow 
\frac{V}{\left( 2\pi \right) ^{3}}\int d\mathbf{p}$, and the distribution $%
\aleph _{m}$ plays the role of the cut-off factor for the integral over $%
p_{3}$. Then, one can conclude that the $p_{\perp },r$ distribution density
of particles produced per unit volume is finite and can be presented as
follows: 
\begin{eqnarray}
&&n_{\mathbf{p}_{\perp },r}^{cr}=\frac{1}{\left( 2\pi \right) ^{3}}%
\int_{-\infty }^{\infty }\Delta N_{m}dp_{3}  \notag \\
&&\,=\frac{1}{\left( 2\pi \right) ^{3}}\left[ e^{-\pi \lambda }\int_{-\left|
qE\right| T/2}^{\left| qE\right| T/2}n_{m}(\beta )dp_{3}+\sqrt{\left|
qE\right| }O(K)\right] ,  \label{dm33} \\
&&n_{m}(\beta )=\left( \tanh \left( E_{m}/2\right) \right) ^{\kappa }. 
\notag
\end{eqnarray}
From (\ref{dm33}), one can estimate the $p_{\perp },r$ distribution density
of the particle production rate, 
\begin{equation}
\frac{dn_{\mathbf{p}_{\perp },r}^{cr}}{dT}=\frac{\left| qE\right| }{\left(
2\pi \right) ^{3}}\left. n_{m}(\beta )\right| _{\pi _{3}=\left| qE\right|
T}e^{-\pi \lambda }.  \label{dm34}
\end{equation}

We have $\left( qET\right) ^{2}\gg M^{2}+\mathbf{p}_{\perp }^{2}$, according
to the condition of stabilization. Then, the high\ and low temperature
limits for the production rate are only defined by the final longitudinal
kinetic momentum $\left| qE\right| T$ and the temperature $\Theta $
relation, $\beta \left| qE\right| T\ll 1$ and $\beta \left| qE\right| T\gg 1$%
, respectively. For simplicity, we assume that $\left| qE\right| T\gg \mu $.
Considering these limits, one obtains for the temperature-dependent term in (%
\ref{dm34}) 
\begin{eqnarray*}
\left. n_{m}(\beta )\right| _{\pi _{3}=\left| qE\right| T} &=&1-2\kappa
e^{-\beta \left| qE\right| T},\;\beta \left| qE\right| T\gg 1, \\
\left. n_{m}(\beta )\right| _{\pi _{3}=\left| qE\right| T} &=&\left[ \beta
\left| qE\right| T/2\right] ^{\kappa },\;\beta \left| qE\right| T\ll 1.
\end{eqnarray*}
We can see that at high temperatures the rate $\frac{dn_{\mathbf{p}_{\perp
},r}^{cr}}{dT}$ is time-dependent; it is much lower than the
zero-temperature value but increases for fermions, and is considerably
higher than the zero temperature value but decreases for bosons.
Consequently, the frequently used notion of a number of particles created
per unit of time makes sense only at low temperatures and in this limit it
coincides with the zero-temperature value of the production rate. We
consider two temperature limits for the $p_{\perp },r$ distribution density (%
\ref{dm33}): low temperatures at $\beta \left( \varepsilon _{\perp }-\mu
\right) \gg 1$, $\varepsilon _{\perp }=\sqrt{M^{2}+\mathbf{p}_{\perp }^{2}}$%
, when all the energies of the particles created in the modes with a given $%
\mathbf{p}_{\perp }$ are considerably higher than the temperature $\Theta $,
and high temperatures, at $\beta \left| qE\right| T\ll 1$, when all the
energies of the created particles are much lower than the temperature $%
\Theta $, 
\begin{eqnarray}
&&n_{\mathbf{p}_{\perp },r}^{cr}=\frac{\sqrt{\left| qE\right| }}{\left( 2\pi
\right) ^{3}}\left[ \sqrt{\left| qE\right| }Te^{-\pi \lambda }+O\left(
K\right) \right] ,\;\kappa =\pm 1,\;\;\beta \left( \varepsilon _{\perp }-\mu
\right) \gg 1,  \notag \\
&&n_{\mathbf{p}_{\perp },r}^{cr}=\frac{\beta \left| qE\right| }{\left( 2\pi
\right) ^{3}}\left[ \left| qE\right| T^{2}/2+O\left( \sqrt{\left| qE\right| }%
T\right) \right] e^{-\pi \lambda },\;\kappa =+1,\;\beta \left| qE\right|
T\ll 1,  \label{dm35} \\
&&n_{\mathbf{p}_{\perp },r}^{cr}=\frac{\sqrt{\left| qE\right| }}{\left( 2\pi
\right) ^{3}}\left[ \frac{4}{\beta \sqrt{\left| qE\right| }}\ln \left( \sqrt{%
\left| qE\right| }T/K\right) e^{-\pi \lambda }+O\left( K\right) \right]
,\;\;\kappa =-1,\;\beta \left| qE\right| T\ll 1.  \notag
\end{eqnarray}
The result at low temperatures is not different from the zero-temperature
result \cite{GavG96a} within the accuracy of our analysis. Integrating
expressions (\ref{dm35}) over $\mathbf{p}_{\perp }$, one finds the total
number of particles created per unit volume at the low-temperature and
high-temperature limits, respectively: 
\begin{eqnarray}
&&\frac{N^{cr}}{V}=J\frac{\left( qE\right) ^{2}T}{\left( 2\pi \right) ^{3}}%
e^{-\pi M^{2}/\left| qE\right| },\;\;\beta \left( M-\mu \right) \gg 1, 
\notag \\
&&\frac{N^{cr}}{V}=\frac{\beta \left| qE\right| ^{3}T^{2}}{\left( 2\pi
\right) ^{3}}e^{-\pi M^{2}/\left| qE\right| },\;\kappa =+1,\;\beta \left|
qE\right| T\ll 1,  \notag \\
&&\frac{N^{cr}}{V}=\frac{\left| qE\right| \ln \left( \sqrt{\left| qE\right| }%
T\right) }{2\pi ^{3}\beta }e^{-\pi M^{2}/\left| qE\right| },\;\;\kappa
=-1,\;\beta \left| qE\right| T\ll 1,  \label{dm36}
\end{eqnarray}
where the summation over $r=\pm 1$ is carried out for the fermions, and only
the leading $T$-dependent terms are presented. From (\ref{dm35}),(\ref{dm36}%
), we can see that the values of the integral mean numbers for fermions at
high temperatures are much lower than the corresponding values at low
temperatures. For bosons, the integral mean numbers at high temperatures are
considerably higher than the corresponding values at low temperatures.

As mentioned in Introduction, thermally-influenced pair production by a
constant electric field has been investigated in several approaches \cite%
{GanKP95,BuhGF80,HalL95,Gie99}. The results are quite contradictory, varying
from the absence of creation to the rates of fermion production higher than
the rate at zero temperature. Now, we are ready to discuss these
contradictions. As has been shown above, the initial thermal distribution
affects the number of states in which pairs are created by a quasi-constant
field. Hence, pair-production exists at any temperatures, and, in
particular, the fermion production rate cannot be higher than the rate at
zero temperature, by any means. Note that our calculations are based on the
generalized Furry representation developed especially for the case of vacuum
instability in accordance with the basic principles of quantum field theory.
On the other hand, all the conclusions of \cite{GanKP95,Gie99} concerning
the pair production rate and/or the mean numbers of pairs created at
non-zero temperatures are based on either the standard real-time, or
imaginary-time, one-loop effective actions. However, such formalisms do not
work in the presence of unstable modes. The real part of the standard
effective action describes the effects of vacuum polarization and has
nothing to do with the time-dependent conduction current of created pairs.
For example, this can be observed at zero temperature (see \cite{Gav06}). In
this case, the information concerning pair-creation comes from the imaginary
part of the standard effective action. The extension of real-time techniques
for finite-temperature quantum electrodynamics with unstable vacuum has been
presented in \cite{GavGF87}. In this article, one can see that the relevant
Green functions in a constant electric field are quite different from the
standard proper-time representation given by Schwinger. Then, the relevant
real-time one-loop effective action must be different from the standard one%
\footnote{%
We will present the relevant real-time one-loop effective action elsewhere.}%
. The standard imaginary-time formalism, obtained under the assumption of a
thermal equilibrium and the appearance of a contradiction with the Pauli
exclusion principle, shows that the attempts of generalization to
far-from-equilibrium systems have failed. The functional Schrödinger picture
used in \cite{HalL95} to calculate the $N^{cr}$ at high temperatures seems
relevant; its asymptotic expressions for $N^{cr}$ at high temperatures are
in agreement with our expressions in (\ref{dm36}).

\subparagraph{\protect\large Acknowledgement}

We thank Yuri Sinyukov for useful discussions, and, in particular, for
calling our attention to Ref. \cite{KhaLT06}.

S.P.G. and J.L.T. thank FAPESP for support. D.M.G. acknowledges the support
of FAPESP and CNPq. S.P.G. is grateful to Universidade Estadual Paulista
(Campus de Guaratinguetá) and Universidade de São Paulo for hospitality.

\appendix

\section*{Appendix}

I. For both the Bose and Fermi cases, the following relations hold: 
\begin{eqnarray}
&&ae^{a^{\dagger }Da}=e^{a^{\dagger }Da}e^{D}a\,,\;\;a^{\dagger
}e^{a^{\dagger }Da}=e^{a^{\dagger }Da}a^{\dagger }e^{-D}\,,  \label{f1} \\
&&e^{a^{\dagger }Da}=\,:\exp \left\{ a^{\dagger }\left( e^{D}-1\right)
a\right\} :\,,  \label{f4}
\end{eqnarray}
where $D$ is a matrix. To prove (\ref{f4}), let us consider the operator
function $F(s)=e^{sa^{\dagger }Da},$ where $s$ is a parameter. The function
is a solution of the following equation: 
\begin{equation*}
\frac{dF(s)}{ds}=a^{\dagger }DaF(s)\,,\;F(0)=1\,.
\end{equation*}
Using relation (\ref{f1}), we can rewrite the right-hand side of the
equation as follows: 
\begin{equation*}
\frac{dF(s)}{ds}=a^{\dagger }F(s)De^{sD}a\,,\,\;F(0)=1\,.
\end{equation*}
Now, we can verify that a solution of such an equation reads 
\begin{equation*}
F(s)=\,:\exp \left\{ a^{\dagger }\left( e^{sD}-1\right) a\right\} :\,.
\end{equation*}
Setting $s=1$, we justify (\ref{f4}).

II. We often use the well-known relation 
\begin{equation}
e^{\lambda a}e^{a^{\dagger }\tilde{\lambda}}=e^{a^{\dagger }\tilde{\lambda}%
}e^{\lambda a}e^{\lambda \tilde{\lambda}}\,,  \label{dm14b}
\end{equation}
where $\lambda $ and $\tilde{\lambda}$ are Grassmann-odd or Grassmann-even
variables depending on statistics. For a product of two normal forms, there
holds a generalization of (\ref{dm14b}), namely, 
\begin{equation}
:e^{a^{\dagger }Da}:\,:e^{a^{\dagger }\tilde{D}a}:\,=\,:e^{a^{\dagger }(D+%
\tilde{D}+D\tilde{D})a}:\,,  \label{dm14c}
\end{equation}
where $D$ and $\tilde{D}$ are some matrices.

III. The projection operator on the vacuum state\ can be written as follows: 
\begin{equation}
P_{0}=|0\rangle \langle 0|=\,:e^{-a^{\dagger }a}:\,.  \label{f3}
\end{equation}
Such a representation was first used by Berezin \cite{Ber65}. One can see
that the operator $P_{0}$ obeys the equations 
\begin{equation*}
aP_{0}=0\,,\;P_{0}a^{\dagger }=0\,,\;P_{0}|0\rangle =|0\rangle \,.
\end{equation*}
Using the Wick theorem, one can see that $:e^{-a^{\dagger }a}:$ is a
solution of these equations.

IV. The trace of a normal product of creation and annihilation operators can
be calculated by using the following path integral representation. Let $%
X\left( a^{\dagger },a\right) $ be an operator expression of creation and
annihilation operators, $a$ and $a^{\dagger }$. Then the trace of its normal
form 
\begin{equation*}
\mathrm{tr}\left\{ :X\left( a^{\dagger },a\right) :\right\}
=\sum_{M=0}^{\infty }\sum_{\{m\}}\left( M!\right) ^{-1}\langle
0|a_{m_{M}}\ldots a_{m_{1}}:X\left( a^{\dagger },a\right)
:a_{m_{1}}^{\dagger }\ldots a_{m_{M}}^{\dagger }|0\rangle \,,
\end{equation*}
can be expressed as the following vacuum mean value: 
\begin{equation}
\mathrm{tr}\left\{ :X\left( a^{\dagger },a\right) :\right\} =\langle
0|T:X\left( a^{\dagger },a\right) :e^{a\left( t_{f}\right) a^{\dagger
}\left( t_{i}\right) }|0\rangle  \label{ap11}
\end{equation}
where the notation $a=a\left( t_{f}\right) $, $a^{\dagger }=a^{\dagger
}\left( t_{i}\right) $ is used for the operators $a$ to the left of $%
:X\left( a^{\dagger },a\right) :$ and $a^{\dagger }$ to the right of $%
:X\left( a^{\dagger },a\right) :$, whereas $T$ is the ordering operator
putting $a\left( t_{f}\right) $ to the left of $:X\left( a^{\dagger
},a\right) :$ and $a^{\dagger }\left( t_{i}\right) $ to the right of $%
:X\left( a^{\dagger },a\right) :$. Using either the Berezin path integral or
the Gaussian integral over ordinary variables, depending on statistics, one
can rewrite (\ref{ap11}) as follows: 
\begin{equation}
\mathrm{tr}\left\{ :X\left( a^{\dagger },a\right) :\right\} =\langle 0|\int
\exp \left\{ \kappa \lambda ^{\ast }\lambda +\lambda ^{\ast }a\right\}
:X\left( a^{\dagger },a\right) :\exp \left\{ a^{\dagger }\lambda \right\}
\Pi d\lambda ^{\ast }d\lambda |0\rangle ,  \label{ap12}
\end{equation}
where $a\left( t_{f}\right) =a$ and $a^{\dagger }\left( t_{i}\right)
=a^{\dagger }$ are used after rewriting.


\begin{thebibliography}{99}
\bibitem{b1} O. Klein, Z. Phys. \textbf{53} (1929) 157; F. Sauter, Z. Phys. 
\textbf{69} (1931) 742; F. Sauter, Z. Phys. \textbf{73} (1931) 547.

\bibitem{S51} J. Schwinger, Phys. Rev. \textbf{82} (1951) 664.

\bibitem{Nikis79} A.I. Nikishov, Sov. Phys. JETP \textbf{30} (1970) 660;
A.I. Nikishov, in \emph{Quantum Electrodynamics of Phenomena in Intense
Fields}, Proc. P.N. Lebedev Phys. Inst. \textbf{111} (Nauka, Moscow, 1979)
153; V.G. Bagrov, D.M. Gitman and Sh.M. Shvartsman, Sov. Phys.-JETP \textbf{%
41} (1975) 191; D.M. Gitman, V.M. Shachmatov and Sh.M. Shvartsman, Sov.
Phys. J. \textbf{4} (1975) 23; V.G. Bagrov, D.M. Gitman and Sh.M.
Shvartsman, Sov. J. Nucl. Phys. \textbf{23} (1976) 394.

\bibitem{GavG96a} S.P. Gavrilov and D.M. Gitman, Phys. Rev. D \textbf{53}
(1996) 7162 [arXiv:hep-th/9603152].

\bibitem{Dre02} A. Ringwald, Phys. Lett. B \textbf{510} (2001) 107;
arXiv:hep-ph/0103185; arXiv:hep-ph/0304139; R. Alkofer, M.B. Hecht, C.D.
Roberts, S.M. Schmidt, and D.V. Vinnik., Phys. Rev. Lett. \textbf{87} (2001)
193902; V.S. Popov, JETP Lett. \textbf{74} (2001) 133; I.M. Dremin, JETP
Lett. \textbf{76} (2001) 185.

\bibitem{NCV99} J.R.S. Nascimento, I. Cho, A. Vilenkin, Phys. Rev. D \textbf{%
60} (1999) 083505 [arXiv:hep-th/9902135].

\bibitem{GMR85} W. Greiner, B. Müller, and J. Rafelski, \emph{Quantum
Electrodynamics of Strong Fields} (Springer-Verlag, Berlin, 1985); W.
Greiner and J. Reinhardt, \emph{Quantum Electrodynamics} (Springer-Verlag,
Berlin, 1994).

\bibitem{Haw75} S.W. Hawking, Commun. Math. Phys. \textbf{43} (1975) 199;
Phys. Rev. D \textbf{14} (1976) 2460.

\bibitem{FroN89} I.D. Novikov and V. P. Frolov, \emph{Physics of black holes}
(Kluwer Academic, Dordrecht, 1989); R.M. Wald, \emph{Quantum field theory in
curved spacetime and black hole thermodynamics} (The University of Chicago
Press, Chicago, 1994).

\bibitem{Grib} A.A. Grib, S.G. Mamaev, and V.M. Mostepanenko, \emph{Vacuum
Quantum Effects in Strong Fields}, (Atomizdat, Moscow, 1988; Friedmann
Laboratory Publishing, St. Petersburg, 1994); N.D. Birrell and P.C.W.
Davies, \emph{Quantum Fields in Curved Space} (Cambridge University Press,
Cambridge 1994); I.L. Buchbinder, S.D. Odintsov and I.L. Shapiro, \emph{%
Effective Action in Quantum Gravity} (IOP Publishing, Bristol and
Philadelphia 1992).

\bibitem{string} N. Seiberg, L. Susskind and N. Toumbas, JHEP \textbf{0006}
(2000) 021; R. Gopakumar, J. Maldacena, S. Minwalla and A. Strominger, JHEP 
\textbf{0006} (2000) 036; P. Mukhopadhyay and A. Sen, JHEP \textbf{0211}
(2002) 047.

\bibitem{NayN05} G.C. Nayak, and P. van Nieuwenhuizen, Phys. Rev. D \textbf{%
71} (2005) 125001 [arXiv:hep-ph/0504070]; G.C. Nayak, Phys. Rev. D \textbf{72%
} (2005) 125010 [arXiv:hep-ph/0510052]; F. Cooper and G.C. Nayak, Phys. Rev.
D \textbf{73} (2006) 065005 [arXiv:hep-ph/0511053].

\bibitem{GelKL06} F.Gelis, K. Kajantie, and T. Lappi, Phys. Rev. Lett. 
\textbf{96} (2006) 032304 [arXiv:hep-ph/0508229].

\bibitem{KhaLT06} D. Kharzeev, E. Levin, and K. Tuchin, arXiv:hep-ph/0602063.

\bibitem{CasNN79} A. Casher, H. Neuberger, and S. Nussinov, Phys. Rev. D 
\textbf{20} (1979) 179; E.G. Gurvich, Phys. Lett. B \textbf{87} (1979) 386.

\bibitem{BatMS77} I.A. Batalin, S.G. Matinyan, and G.K. Savvidy, Sov. J.
Nucl. Phys. \textbf{26} (1977) 214; J. Ambjorn and R. Hughes, Phys. Lett. B 
\textbf{113} (1982) 305; Ann. Phys. (N.Y.) \textbf{145} (1983) 340.

\bibitem{YilC80} A. Yildiz and P. H. Cox, Phys. Rev. D \textbf{21} (1980)
1095; M. Claudson, A. Yildiz and P. H. Cox, Phys. Rev. D \textbf{22} (1980)
2022.

\bibitem{ManS01} J. Manjavidze and A. Sissakian, Phys. Rept. \textbf{346}
(2001) 1; D.D. Dietrich, G.C. Nayak, and W. Greiner, Phys. Rev. D \textbf{64}
(2001) 074006; J.D. de Deus, E.G. Ferreiro, C. Pajares, and R. Ugoccioni,
Phys. Lett. B \textbf{581} (2004) 156.

\bibitem{McLG05} L. McLerran, and M. Gyulassy, Nucl. Phys. A \textbf{750}
(2005) 30.

\bibitem{Schu04} Y. Schutz, J. Phys. G \textbf{30} (2004) S903.

\bibitem{McLV94} L.V. Gribov, E.M. Levin, and M.G. Ryskin, Phys. Rep. 
\textbf{100} (1983) 1; L. McLerran, and R. Venugopalan, Phys. Rev. D \textbf{%
49} (1994) 2233, 3352; \textbf{50} (1994) 2225; \textbf{59} (1999) 094002;
A. Ayala, J. Jalilian-Marian, L. McLerran, and R. Venugopalan, Phys. Rev. D 
\textbf{53} (1996) 458; A.H. Mueller, and J. Qiu, Nucl. Phys. B \textbf{268}
(1986) 427.

\bibitem{IanV03} L. McLerran, Lectures given at the 40'th Schladming Winter
School: Dense Matter, March 3-10 2001, arXiv:hep-ph/0104285; E. Iancu, A.
Leonidov, L. McLerran, Lectures given at Cargese, France, 6-19 Aug 2001,
arXiv:hep-ph/0202270; E.Iancu, R. Venugopalan, Quark Gluon Plasma 3, Eds.
R.C. Hwa and X.N. Wang, World Scientific [arXiv:hep-ph/0303204]; H. Weigert,
Prog. Part. Nucl. Phys. \textbf{55} (2005) 461 [arXiv:hep-ph/0501087].

\bibitem{KovMW95} A. Kovner, L. McLerran,, and H. Weigert, Phys. Rev. D 
\textbf{52} (1995) 6231.

\bibitem{BuhGF80} I.L. Bukhbinder, D.M. Gitman, and V.P. Frolov, Izv. Vuzov
Fizika \textbf{23}, N.6 (1980) 77 [English. transl.: Sov. Phys. J. \textbf{23%
} (1980) 529].

\bibitem{HalL95} J. Hallin and P. Liljenberg, Phys. Rev. D \textbf{52}
(1995) 1150.

\bibitem{GanKP95} A.K. Ganguly, P.K. Kaw, and J.C. Parikh, Phys. Rev. C 
\textbf{51} (1995) 2091.

\bibitem{Gie99} P.H. Cox, W.S. Hellman, and A. Yildiz, Ann. Phys. (N.Y) 
\textbf{154} (1984) 211; M. Loewe and J.C. Rojas, Phys. Rev. D \textbf{46}
(1992) 2689; P. Elmfors and B.-S. Skagerstam, Phys. Lett. B \textbf{348}
(1995) 141; \textbf{376} (1996) 330(E); A.K. Ganguly, arXiv:hep-th/9804134;
H. Gies, Phys. Rev. D \textbf{60} (1999) 105002.

\bibitem{Gitman} D.M. Gitman, J. Phys. A \textbf{10} (1977) 2007; E.S.
Fradkin, D.M. Gitman and S.M. Shvartsman, \emph{Quantum Electrodynamics with
Unstable Vacuum} (Springer-Verlag, Berlin 1991)

\bibitem{BFG81} I.L. Buchbinder and D.M. Gitman, Izv. V. Fiz. (Sov. Phys.
J.) 3 (1979) 90 ; ibid 4 (1979) 55; ibid 7 (1979) 16; I.L. Buchbinder, E.S.
Fradkin and D.M. Gitman, Fortschr. Phys. \textbf{29} (1981) 187.

\bibitem{GavG91} S.P. Gavrilov and D.M. Gitman, Report MIT, Massachusetts,
1991, CTP\#1995 (unpublished); S.P. Gavrilov, Russian Phys. J. \textbf{35},
No. 7 (1992) 652; ibid, \textbf{35}, No. 10 (1992) 969; ibid, \textbf{36},
No. 3 (1993) 269; S.P. Gavrilov and D.M. Gitman, ibid, \textbf{36}, No. 5
(1993) 448; Izv. Vuzov. Fiz. No. 4 (1995) 83.

\bibitem{AmbHN83} J. Ambjorn, and R. J. Hughes, and N.K. Nielsen Ann. Phys.
(N.Y.) \textbf{150} (1983) 92.

\bibitem{GavGF87} S.P. Gavrilov, D.M. Gitman, and E.S. Fradkin, Yad. Fiz. 
\textbf{46} (1987) 172 [English. transl.: Sov. J. Nucl. Phys. (USA) \textbf{%
46} (1987) 107].

\bibitem{Ber65} F. A. Berezin, \emph{The method of second quantization}
(Nauka, Moscow, 1965) [English transl.: Academic Press, New York (1966)].

\bibitem{Thermal} D.M. Gitman and V.P. Frolov, Yad Fiz. \textbf{28} (1978)
552; V.P. Frolov and D.M. Gitman, J. Phys. A. \textbf{15} (1978) 1329.

\bibitem{GleM84} N. Glendenning and T. Matsui, Phys. Lett. B \textbf{141}
(1984) 419; G. Gatoff, A.K.\ Kerman, and T. Matsui, Phys. Rev. D\textbf{\ 36}
(1987) 114; A. Bialas, W. Czy, A. Dyrek, and W. Florkowski, Nucl. Phys. B 
\textbf{\ 296} (1988) 611.

\bibitem{KluME98} Y. Kluger, E. Mottola, and J.M. Eisenberg, Phys. Rev. D 
\textbf{58} (1998) 125015.

\bibitem{Dunn04} G.V. Dunne, \emph{Heisenberg-Euler effective Lagrangians:
Basics and extensions}, in Ian Kogan Memorial Volume, \emph{From fields to
strings: Circumnavigating theoretical physics}, Eds. M Shifman, A.
Vainshtein and J. Wheater, World Scientific [arXiv:hep-th/0406216].

\bibitem{ItzZub80} C. Itzykson and J.-B. Zuber, \emph{Quantum Field Theory}
(Mc-Graw Hill, New York,1980).

\bibitem{KhaT05} D. Kharzeev and K. Tuchin, Nucl. Phys. A 753 (2005) 316
[arXiv:hep-ph/0602063].

\bibitem{GavGO97} S.P. Gavrilov, D.M. Gitman, and S.D. Odintsov, Int. J.
Mod. Phys. A \textbf{12} (1997) 4837; S.P. Gavrilov, D.M. Gitman, and A.E.
Gonçalves, Int. J. Mod. Phys. A \textbf{16} (2001) 4235.

\bibitem{NarN70} N.B. Narozhny and A.I. Nikishov, Sov. J. Nucl. Phys. (USA) 
\textbf{11} (1970) 596.

\bibitem{Gav06} S.P. Gavrilov, J. Phys. A \textbf{39} (2006) 6407
[arXiv:hep-th/0510093].

\bibitem{Rindler} S.A. Fulling, Phys. Rev. D \textbf{7} (1973) 2850; P.C.W.
Davies, J. Phys. A \textbf{8} (1975) 609; W.G. Unruh, Phys. Rev. D \textbf{14%
} (1976) 870.

\bibitem{KluES93} Y. Kluger, J.M. Eisenberg, and B. Svetitsky, Int. J. Mod.
Phys. E \textbf{2} (1993) 333; F. Cooper, J.M. Eisenberg, Y. Kluger, E.
Mottola, and B. Svetitsky, Phys. Rev. D \textbf{48} (1993) 190; J.C.R. Bloch 
\emph{et al}, Phys. Rev. D \textbf{60} (1999) 116011; D.V.Vinnik \emph{et al}%
, Eur. Phys. J. C \textbf{22} (2001) 341 [arXiv:nucl-th/0103073].

\bibitem{Sch98} S.M. Schmidt \emph{et al}., Int. J. Mod. Phys. E \textbf{7}
(1998) 709; R.S. Bhalerao and G.C. Nayak, Phys. Rev. C \textbf{61} (2000)
054907.

\bibitem{LapML06} T. Lappi and L. McLerran, hep-ph/0602189; Nucl. Phys. A 
\textbf{772} (2006) 200.

\bibitem{NieO78} N.K. Nielsen and P. Olesen, Nucl. Phys. B \textbf{144}
(1978) 376.

\bibitem{AmbO80} J. Ambjorn and P. Olesen, Nucl. Phys. B \textbf{170} (1980)
60.

\bibitem{Flo83} C.A. Flory, Phys. Rev. D \textbf{28} (1983) 1425; Y.M. Cho,
J.H. Kim, and D.G. Pak, Mod. Phys. Lett. A \textbf{21} (2006) 2789.

\bibitem{Spo82} B.L. Spokoiny, Sov. J. Nucl. Phys. \textbf{36} (1982) 277;
Phys. Lett. A \textbf{88} (1982) 328 .

\bibitem{Bia99} A.\ Bialas, Phys. Lett. B \textbf{466} (1999) 301.
\end{thebibliography}
\end{document}